\documentclass[%
 amsmath,amssymb,
 aps,
]{revtex4-2}

\usepackage{graphicx}
\usepackage{dcolumn}
\usepackage{amsmath,amssymb,bm,bbm}
\usepackage{xcolor}
\usepackage{hyperref}

\usepackage{ragged2e}

\usepackage[T1]{fontenc}
\usepackage[utf8]{inputenc}
\usepackage{lmodern} 

\begin{document}

\preprint{APS/123-QED}

\title{Efficient simulation framework for modeling collective emission in ensembles of inhomogeneous solid-state emitters}
\author{Qingyi Zhou*}
\author{Wenxin Wu}
\author{Maryam Zahedian}
\author{Zongfu Yu}
\author{Jennifer T. Choy*}

\affiliation{Department of Electrical and Computer Engineering, University of Wisconsin--Madison, Madison, WI 53706, USA}


\begin{abstract}
An efficient simulation framework is proposed to model collective emission in disordered ensembles of quantum emitters. 
Using a cumulant expansion approach, the computational complexity scales polynomially as opposed to exponentially with the number of emitters, enabling Monte Carlo sampling over a large number of realizations. 
The framework is applied to model negatively charged silicon-vacancy (SiV$^{-}$) centers inside diamond. 
Incorporating spatial disorder and inhomogeneous broadening, we obtain statistically averaged responses over hundreds of SiV$^{-}$ clusters. 
These simulations reveal two signatures of collective behavior. 
First, dynamics of fully inverted clusters show that superradiant emission occurs only with sufficiently large emitter number and high quantum efficiency. Unlike ideal Dicke superradiance, the burst is substantially suppressed by strong near-field dipole-dipole interaction, consistent with existing theoretical predictions. 
Second, under continuous-wave excitation we compute photoluminescence-excitation spectra, which exhibit interaction-induced broadening in the distribution of resonance peaks. 
The corresponding density of states also displays a non-zero skewness. 
Overall, by incorporating realistic inhomogeneities in emitter clusters, our framework is able to predict statistics for disordered ensembles that can be compared to experiments directly. 
Our approach generalizes to other types of emitters, including atoms, molecules, and quantum dots, thus providing a practical tool for analyzing collective behavior in realistic quantum systems. 
\end{abstract}

\maketitle


\section{Introduction}
Photon-mediated collective phenomena between quantum emitters have been a foundational building block for quantum information science \cite{Sipahigil2016, Tamara2021, Garcia2017a, Dolde2014, Sheremet2023} as well as quantum sensing applications \cite{Hosten2016, Guillaume2023}. 
Understanding such collective behavior is therefore essential from both fundamental science and application perspectives. 
While numerous demonstrations of collective phenomena have been performed with optically trapped atoms \cite{Guerin2016_PRL, Yan2023_PRL, Browaeys2020_NatPhys}, the interatomic spacing is ultimately constrained by the diffraction limit of the trapping optics - bounded to $\lambda/2$ in optical lattices - which limits the parameter space, integrability, and control of this interaction \cite{Sortais2007_PRA, Wang2018_PRL}.

Alternatively, color centers in diamond have attracted much attention over the past decade since they present atom-like emitter characteristics in a solid-state environment, with narrow optical linewidths, high measurement sensitivity, as well as compatibility with on-chip photonic integration \cite{Hepp2014, Pingault2017, Schirhagl2014, Bhaskar2017}. Crucially, methods to deterministically position color centers in diamond, through nitrogen delta-doping \cite{Ohno2012_APL}, implantation with focused ion beam \cite{Pezzagna2011_PSSA, xu2023fabrication}, and ultrafast multi-photon (laser writing) processes \cite{Chen2017_NatPhoton, Chen2019_Optica}, have resulted in deeply subwavelength ($\sim50$ nm) positioning accuracy \cite{Pezzagna2011_PSSA}.
Although the emission properties of individual color centers have been well understood both theoretically and experimentally \cite{Hepp2014, Rogers2014a, Hepp2014Thesis}, studies of photon-mediated interactions between multiple emitters remain relatively limited \cite{Sipahigil2016, Evans2018, Day2022}. Experimental works have begun to observe collective emission in clusters with $N>2$ emitters \cite{Angerer2018, Pallmann2024}, yet a comprehensive and efficient simulation framework that captures realistic disordered clusters remains lacking. 
Modeling realistic ensembles faces two major challenges. 
First, the interacting quantum emitters form a many-body system, whose Hilbert space grows exponentially with the number of emitters $N$. Simulating the cooperative dynamics for such a many-body system in the multi-excitation regime is challenging due to the high computational cost. 
Second, in experiments these emitters exhibit intrinsic disorder in position and orientation, which breaks the spatial symmetries assumed in idealized models, thus making the system drastically different contrary to a highly-symmetric configuration (e.g., a ring or an infinite array of identical emitters) \cite{Masson2022, Garcia2017a}. 
Furthermore, the presence of structured photonic environments (e.g., the air–diamond interface) not only modifies the emission process of a single quantum emitter \cite{Zahedian2023, Wambold2020, Chakravarthi2020, Kim2025}, but also the resonant dipole-dipole interaction (RDDI) \cite{Dung2002, Bay1997, Ying2019, Gangaraj2022}, thus further invalidating simplifications based on symmetric collective states. 
Together, these difficulties have created a huge gap between theoretical descriptions and experimental reality, making it difficult to interpret measurements, let alone engineer collective effects for quantum applications \cite{Torcal2024}. 

In this paper, we address the aforementioned challenges by developing an efficient simulation framework that incorporates realistic non-idealities. 
Our approach is based on the cumulant expansion approach \cite{Kubo1962, Plankensteiner2022}, which has been validated for capturing superradiant effects in atomic arrays \cite{Robicheaux2021, Bigorda2023, Ostermann2024}. 
With this technique, we aim to understand the emission property of color-center ensembles. 
More specifically, our framework takes into account the spatial distribution of color centers, inhomogeneous broadening of transition frequencies, as well as realistic excitation and detection channels through resonant photoluminescence spectroscopy and fluorescence decay measurements.
By truncating the Heisenberg equations to the second order, the computational complexity is reduced significantly, enabling large-scale Monte Carlo sampling over hundreds of cluster configurations. 
This powerful framework allows us to uncover key signatures of their collective behavior in both time domain and frequency domain. 
In the time domain, we have understood how the number of emitters affects the time evolution of photon emission. 
We identify the threshold conditions for superradiance, specifying that both sufficient emitter number and high quantum efficiency are necessary to observe a superradiant burst. 
In the frequency domain, we show for photoluminescence excitation (PLE) experiments that dipole-dipole interactions induce a significant broadening of the distribution of resonance peak positions. 
By capturing the full complexity of clusters encountered in realistic experiments, our simulations are able to predict statistically averaged signatures that can be directly compared to measurements. 
This framework thus provides a direct pathway to quantifying the role of dipole–dipole interactions within color center ensembles. 

The rest of this paper is organized as follows: in Sec. II we derive equations for emission dynamics from the Hamiltonian of a multi-emitter system that takes into account both inhomogeneity and photon-mediated dipole-dipole interaction. In Sec. III we present the results obtained by applying our simulation technique to the system of interest, which is a cluster of SiV$^{-}$ centers. 
To capture random variations between different clusters, we perform Monte Carlo simulations over hundreds of disordered clusters, enabling a statistical prediction of collective emission properties. 
More specifically, in the first subsection of Sec. III we specify the model parameters, quantifying the extent of disorder. 
In the second subsection of Sec. III, we analyze the decay process of a fully excited SiV$^{-}$ ensemble, identifying the conditions under which superradiance remains observable. 
In the third subsection we show that under continuous-wave (CW) excitation, RDDI leads to differences in PLE statistics beyond inhomogeneous broadening. 
Finally, we conclude in Sec. IV that dipole-dipole interaction inside SiV$^{-}$ cluster does lead to observable effects, and should be possible to verify by comparing statistically averaged responses of SiV$^{-}$ clusters with different densities. 

\begin{figure}
  \centering
  \includegraphics[width=0.85\linewidth]{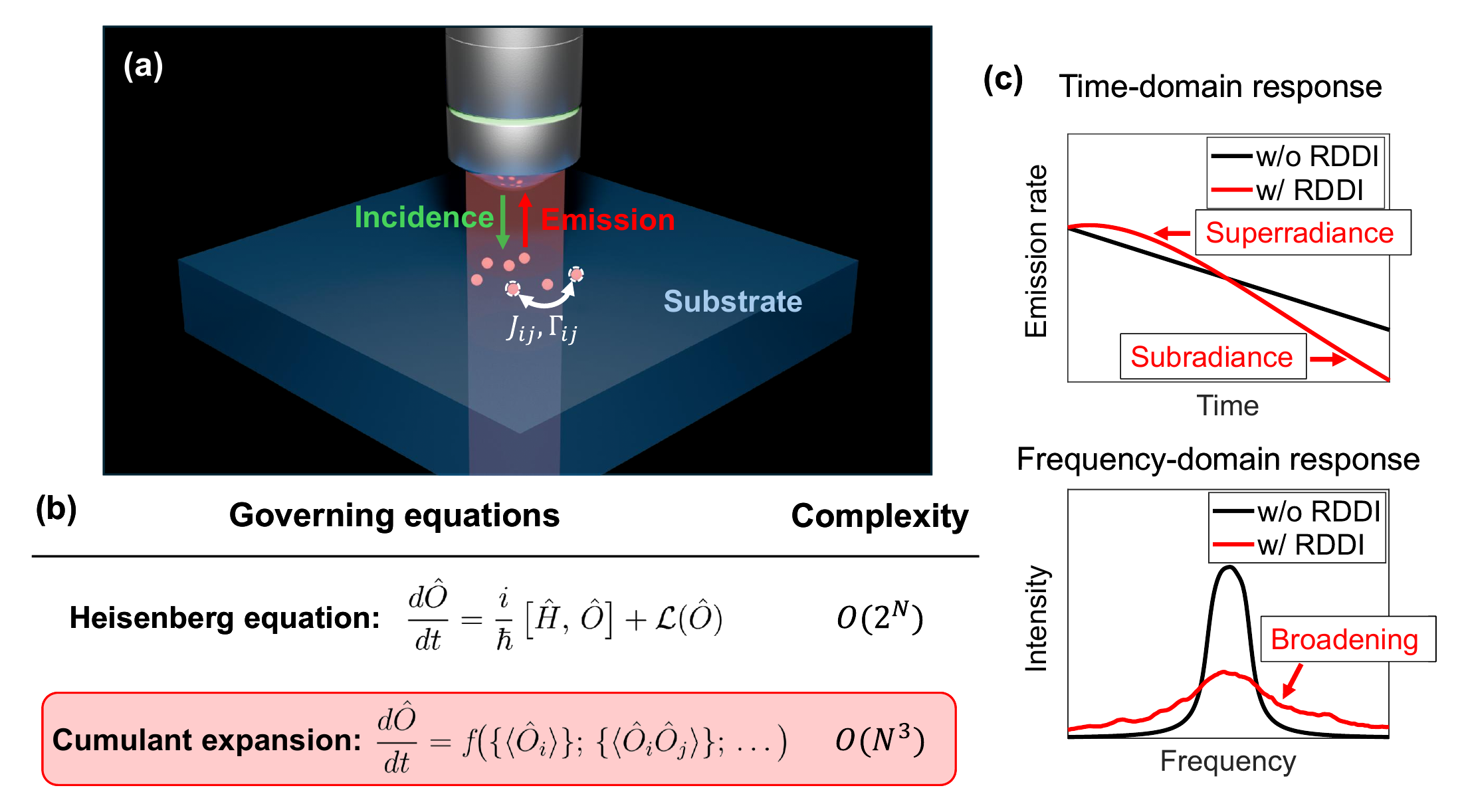}
  \caption{Schematic illustration of the theoretical approach used in this work. 
  (a) High-level sketch of the considered model. Multiple quantum emitters are randomly positioned within the photonic environment. The emitters are driven by an external laser; the emitted light is collected and analyzed. The photon-mediated dipole-dipole interactions, represented by $J_{ij}$ and $\Gamma_{ij}$, lead to nontrivial collective behavior. 
  (b) When using the full set of Heisenberg equations to simulate $N$ emitters, the complexity scales exponentially. The computational cost can be reduced by truncating the equations of motion via a cumulant expansion approach. 
  (c) The developed framework predicts both time-domain dynamics and spectral responses of emitter ensembles. It captures several important features arising from dipole–dipole interactions, including superradiance, subradiance, and interaction-induced broadening. }
  \label{fig:fig1}
\end{figure}

\section{Model \& Methods}
Without loss of generality, our model consists of an ensemble of $N$ quantum emitters, as shown in Figure 1a. 
Each quantum emitter is modeled as an ideal two-level system (TLS), with ground state $|1\rangle$ and excited state $|2\rangle$. For the $i$-th emitter, the transition frequency is denoted as $\omega_{0}+\delta_{i}$, where $\delta_{i}$ captures static inhomogeneous detuning. 
The emitters are driven by a CW laser at frequency $\omega_{d}$. 
While the system is coupled to a photon bath, the full Hamiltonian can be simplified by applying the Born-Markov approximation and eliminating the photon degrees of freedom \cite{Carmichael1993, Ficek2004}. For the $i$-th emitter we define the corresponding operators $\hat{\sigma}_{i}^{21} = |2_{i}\rangle \langle 1_{i} |$, $\hat{\sigma}_{i}^{12} = |1_{i}\rangle \langle 2_{i} |$, and $\hat{\sigma}_{i}^{22} = |2_{i}\rangle \langle 2_{i} |$. The effective Hamiltonian takes the form \cite{Ficek2002}
\begin{align}
    \hat{H}_\text{eff} &= \hat{H}_\text{free} + \hat{H}_\text{couple} + \hat{H}_\text{drive} \nonumber \\
    & = \hbar \sum_{i} (\omega_{0} + \delta_{i} - \omega_{d}) \hat{\sigma}_{i}^{22} 
    + \hbar \sum_{j\neq i} J_{ij} \hat{\sigma}_{i}^{21} \hat{\sigma}_{j}^{12} 
    + \frac{\hbar}{2} \sum_{i} (\Omega_{i} \hat{\sigma}_{i}^{21} + h.c.),
    \label{eq:eff_Hamiltonian}
\end{align}
where $J_{ij}$ represents the coherent part of dipole-dipole interaction between the $i$-th and $j$-th TLSs. The laser driving term $\hat{H}_\text{drive}$ includes the complex Rabi frequency $\Omega_{i}=\frac{2}{\hbar} \bm{\mu}_{i} \cdot \bm{E}_\text{inc}(\bm{r}_{i})$, where $\bm{\mu}_{i}$ denotes the transition dipole moment of the $i$-th emitter, and $h.c.$ stands for Hermitian conjugate. The above equation has already been transformed into the $e^{-i\omega_{d} t}$ rotating frame. 

In order to solve for the dynamics of this system, we work in the Heisenberg picture and apply the cumulant expansion approach, which has been introduced to simulate the superradiance effect \cite{Robicheaux2021, Bigorda2023}. The equation of motion for an arbitrary operator $\hat{O}$ can be derived as
\begin{equation}
    \frac{d\hat{O}}{dt} = \frac{i}{\hbar} [\hat{H}_\text{eff}, \hat{O}] + \mathcal{L}(\hat{O}), 
    \label{eq:Heisenberg}
\end{equation}
where the superoperator $\mathcal{L}(\hat{O})$ represents the Lindblad term which takes into account the dissipative part of dipole-dipole interaction \cite{Carmichael1993}: 
\begin{equation}
    \mathcal{L}(\hat{O}) = \sum_{i, j} \Gamma_{ij} \left[ \hat{\sigma}_{i}^{21} \hat{O} \hat{\sigma}_{j}^{12} 
    - \frac{1}{2} \hat{\sigma}_{i}^{21} \hat{\sigma}_{j}^{12} \hat{O} 
    - \frac{1}{2} \hat{O} \hat{\sigma}_{i}^{21} \hat{\sigma}_{j}^{12} \right].
    \label{eq:Lindblad}
\end{equation}
In the above expressions, the RDDI contains a coherent part $J_{ij}$ as well as a dissipative part $\Gamma_{ij}$ \cite{Ficek2004}. For an arbitrary photonic environment, these can be obtained from the dyadic Green's function between the $i$-th and $j$-th emitters \cite{Garcia2017a, Garcia2017b, Gangaraj2022, Zhou2024}: 
\begin{equation}
    J_{ij} - i\frac{\Gamma_{ij}}{2} = -\frac{\omega_{0}^{2}}{\hbar \epsilon_{0} c_{0}^2} \bm{\mu}_{i} \cdot \mathbf{G}(\bm{r}_{j}, \bm{r}_{i}) \cdot \bm{\mu}_{j}, 
\end{equation}
where $\bm{\mu}_{i}$ and $\bm{\mu}_{j}$ correspond to the transition dipole moment of the $i$-th and $j$-th emitter, while $\mathbf{G}(\bm{r}_{j}, \bm{r}_{i})$ gives the electric field at position $\bm{r}_{j}$ induced by a dipole at $\bm{r}_{i}$. We include the air-substrate interface throughout this paper, which makes the photonic environment inhomogeneous. The detailed derivation for the dyadic Green's function $\mathbf{G}(\bm{r}_{j}, \bm{r}_{i})$ is provided in Supplementary Note S1. 

While the number of operator expectation values $\langle\hat{O}\rangle \triangleq \text{Tr}[\hat{\rho}\hat{O}]$ appearing in Equation~(\ref{eq:Heisenberg}) grows exponentially with the number of emitters $N$, it is possible to truncate these equations, by approximating averages of higher-order operators with combinations of lower-order ones \cite{Bigorda2023}. More specifically, in this work we apply the second-order cumulant expansion by replacing the third-order moments with 
\begin{equation}
    \langle \hat{O}_{1} \hat{O}_{2} \hat{O}_{3} \rangle 
\approx \langle \hat{O}_{1} \rangle \langle \hat{O}_{2} \hat{O}_{3} \rangle 
+ \langle \hat{O}_{2} \rangle \langle \hat{O}_{1} \hat{O}_{3} \rangle 
+ \langle \hat{O}_{3} \rangle \langle \hat{O}_{1} \hat{O}_{2} \rangle 
- 2 \langle \hat{O}_{1} \rangle \langle \hat{O}_{2} \rangle \langle \hat{O}_{3} \rangle,
\end{equation}
which helps reduce the number of coupled equations to scale as $O(N^{2})$. The general implementation of this truncation in open quantum emitter ensembles has been discussed in the literature \cite{Bigorda2023}. 
Such cumulant expansion approach has been successfully applied to solving many-body systems in various fields \cite{Kubo1962, Bigorda2023, Piper2023}, motivating us to use this method in combination with Monte Carlo techniques to model clusters of color centers with inhomogeneity. 

Under the second-order cumulant expansion framework, we obtain a closed set of coupled ordinary differential equations (ODEs). The equations are derived with the help of the ``QuantumCumulants.jl'' framework \cite{Plankensteiner2022}. For brevity, we present in the main text only the ODEs for first-order operators:
\begin{align}
\frac{d}{dt} \langle \hat{\sigma}^{12}_i \rangle &=
\left[-i (\omega_{0} + \delta_i - \omega_{d}) - \frac{\Gamma_{ii} + \Gamma^\text{nrad}_{i}}{2} \right] \langle \hat{\sigma}^{12}_i \rangle 
+ {i\Omega_i} \left( \langle \hat{\sigma}^{22}_i \rangle - \frac{1}{2} \right) \nonumber \\ 
&+ \sum_{j \neq i} \left(i J_{ij} + \frac{\Gamma_{ij}}{2} \right) \cdot \left( 2\langle \hat{\sigma}^{12}_j \hat{\sigma}^{22}_i \rangle - \langle \hat{\sigma}^{12}_j \rangle \right), 
\end{align}
\begin{align}
\frac{d}{dt} \langle \hat{\sigma}^{22}_{i} \rangle = 
- (\Gamma_{ii} + \Gamma^\text{nrad}_{i}) \langle \hat{\sigma}^{22}_{i} \rangle + \sum_{j \neq i} \left\{ \left(iJ_{ji} - \frac{\Gamma_{ji}}{2}\right) \langle \hat{\sigma}^{21}_{j} \hat{\sigma}^{12}_{i} \rangle 
+ h.c. \right\}  
- \frac{1}{2} \left( i\Omega_i \langle \hat{\sigma}^{21}_{i} \rangle + h.c. \right). 
\label{eq:first_order}
\end{align}
The full set of equations including all second-order operators is provided in Supplementary Note S2. Notice that here the non-radiative decay rate $\Gamma^\text{nrad}_{i}$ has been considered for the $i$-th emitter. To validate the correctness of our implementation, we benchmark against established results for superradiant cooperative dynamics obtained via the same cumulant expansion strategy \cite{Bigorda2023} (details can be found in Supplementary Note S3). While the number of variables involved scales as $O(N^{2})$, each equation still contains a summation over all $N$ emitters at each time-step. Therefore, the computational complexity for solving the equation scales as $O(N^{3})$, as summarized in Figure 1b. 

The above model has already taken into account dipole-dipole interaction between emitters, as well as inhomogeneity. 
The model is not restricted to a certain type of quantum emitter: beside color centers, any ensemble of two-level emitters such as atoms, molecules, or quantum dots can be treated within the same formalism. 
This feature ensures the broad applicability of our framework and its numerical implementation. 
In the next section, we apply this technique to spatially disordered emitter clusters and show that the simulation framework can predict both their time-domain and frequency-domain responses, as shown by the two panels in Figure 1c. 
Based on the numerical simulation results, we demonstrate that the RDDI remains strong enough to produce effects that are experimentally detectable, even in the presence of inhomogeneous broadening. 

\section{Results}
\begin{figure}
  \centering
  \includegraphics[width=1.0\linewidth]{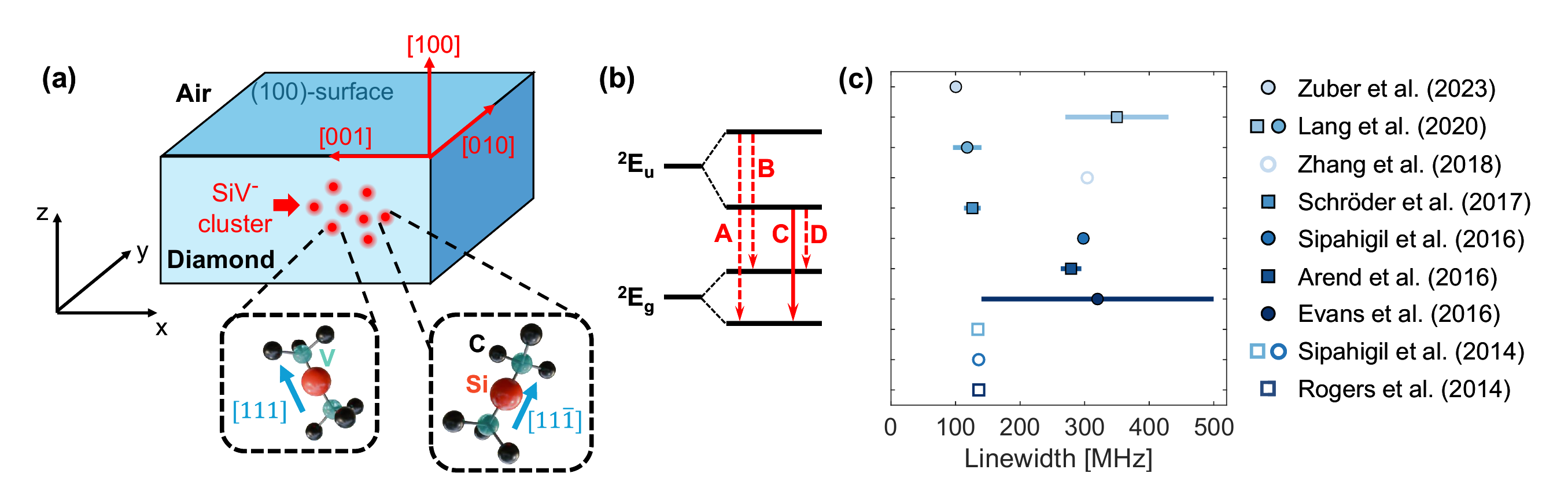}
  \caption{The key properties of the SiV$^{-}$ ensembles modeled in this paper. 
  (a) Schematic illustration of an ensemble of SiV$^{-}$ centers embedded in diamond. The SiV$^{-}$ centers have different orientations, as indicated by the insets.
  (b) Level structure of a single SiV$^{-}$ center. Although four optical transitions are allowed (marked as A, B, C, D), we only focus on the C transition and model each SiV$^{-}$ center as an ideal two-level system. 
  (c) Experimental linewidth measurements compiled from the literature \cite{Rogers2014b, Sipahigil2014, Evans2016, Arend2016, Sipahigil2016, Schroder2017, Zhang2018, Lang2020, Zuber2023}. Each filled marker represents the reported mean value, while the horizontal line denotes the corresponding error bar. If no error bar has been reported, an unfilled marker is used. 
  }
  \label{fig:fig2}
\end{figure}
\subsection{Simulation setup}
In this section we consider an ensemble of $N$ SiV$^{-}$ color centers in diamond, as shown in Figure 2a. Each SiV$^{-}$ center consists of an interstitial silicon atom and a split vacancy. The electronic structure of the SiV$^{-}$ center is shown in Figure 2b. 
Spin-orbit coupling lifts the degeneracy of both the ground and the excited manifolds, producing a ground-state doublet split by $\sim 50$ GHz and an excited state doublet split by $\sim 250$ GHz \cite{Hepp2014}. 
The resulting fine structure gives rise to four dipole-allowed optical transitions, labeled as A, B, C, and D, as shown in Figure 2b. 
At cryogenic temperatures below 4 K, these lines are well resolved, since the smallest separation between them still greatly exceeds typical inhomogeneous broadening of individual color centers. 
We therefore restrict ourselves to a single line, and focus on the C transition centered at $\omega_{0}=2\pi\times 406.7001$ THz \cite{Sipahigil2014}. Each SiV$^{-}$ is then modeled as an ideal TLS. 

Many theoretical studies related to cooperative emission begin with highly regular emitter geometries, such as arrays or chains \cite{Garcia2017a, Masson2020, Masson2022}, because symmetry simplifies the analysis and enables tractable models of Dicke superradiance. 
In contrast, color centers produced by ion implantation form disordered clusters: their spatial positions and orientations vary from one realization to another \cite{Pezzagna2011_PSSA, xu2023fabrication}. Similar variability is also present for thin layers formed by delta doping \cite{Ohno2012_APL}, where emitters are confined in depth but remain laterally disordered. 
Due to such sample-to-sample variability and inhomogeneous broadening, it is difficult to match experimental data with simulations of any specific cluster. 
To overcome this issue we apply statistical analysis by performing Monte Carlo sampling over hundreds of independent cluster realizations. Similar approach has already been utilized to understand the collective behavior of few-atom systems \cite{Damanet2016}. We sample positions, orientations, and resonance frequencies from the given distributions, then compute ensemble-averaged observables such as photon emission rate. 
By analyzing these simulation results, signatures of collective behavior that are robust to disorder can be extracted. 

We now introduce the details of our Monte Carlo simulation setup. 
Each realization consists of one cluster containing $N$ SiV$^{-}$ centers, whose positions $\bm{r}_{i}$ are distributed randomly. The lateral coordinates $(x, y)$ are sampled uniformly within a circular region of diameter $60$ nm in the $xy$-plane, which approximates the ion implantation area achieved in our experiments. 
The depth in $z$ direction follows a Gaussian distribution \cite{Hepp2014Thesis}. By assuming that our SiV$^{-}$ clusters are fabricated using ion beam implantation, the mean and standard deviation of this depth distribution can be estimated with the help of SRIM simulations (details are provided in Supplementary Note S4). 
As for the dipole moment orientations, we choose to use a diamond with $(100)$-surface cut which is more commonly accessible. The transition dipole orientation $\bm{\mu}_{i}$ is parallel to the three-fold symmetry axis \cite{Hepp2014, Rogers2014a, Hepp2014Thesis}, which is chosen randomly from four equivalent directions $[111]$, $[1\bar{1}\bar{1}]$, $[\bar{1}\bar{1}1]$, and $[\bar{1}1\bar{1}]$ (details can be found in Supplementary Note S5). 
Due to inhomogeneous broadening, the individual resonance frequencies $\omega_{0} + \delta_{i}$ must also be sampled randomly. Based on measurements reported in the literature \cite{Sipahigil2014, Zuber2023}, we sample the spectral disorder from a zero-mean Gaussian distribution $\delta_{i} \sim \mathcal{N}(0, 2\pi \times 400 \text{MHz})$, reflecting the typical inhomogeneous spread for low-strain, high-purity SiV$^{-}$ ensembles. 
To determine a physically reasonable linewidth, Figure 2c compiles the linewidth measurement results reported in the literature \cite{Rogers2014b, Sipahigil2014, Evans2016, Arend2016, Sipahigil2016, Schroder2017, Zhang2018, Lang2020, Zuber2023}. Several studies have reported linewidths close to $100$ MHz at $T=4$K, close to the lifetime limit. Accordingly, we set the total decay rate used in our simulations as $115$ MHz. 
We fix the refractive index of diamond at $n=2.4$ when calculating the dyadic Green's function, since diamond’s dispersion is negligible around the SiV$^{-}$ zero-phonon line at $737$ nm \cite{Phillip1964}. The above parameter distributions will be used throughout the paper. 

\subsection{Time-evolution results}

\begin{figure}
  \centering
  \includegraphics[width=1.0\linewidth]{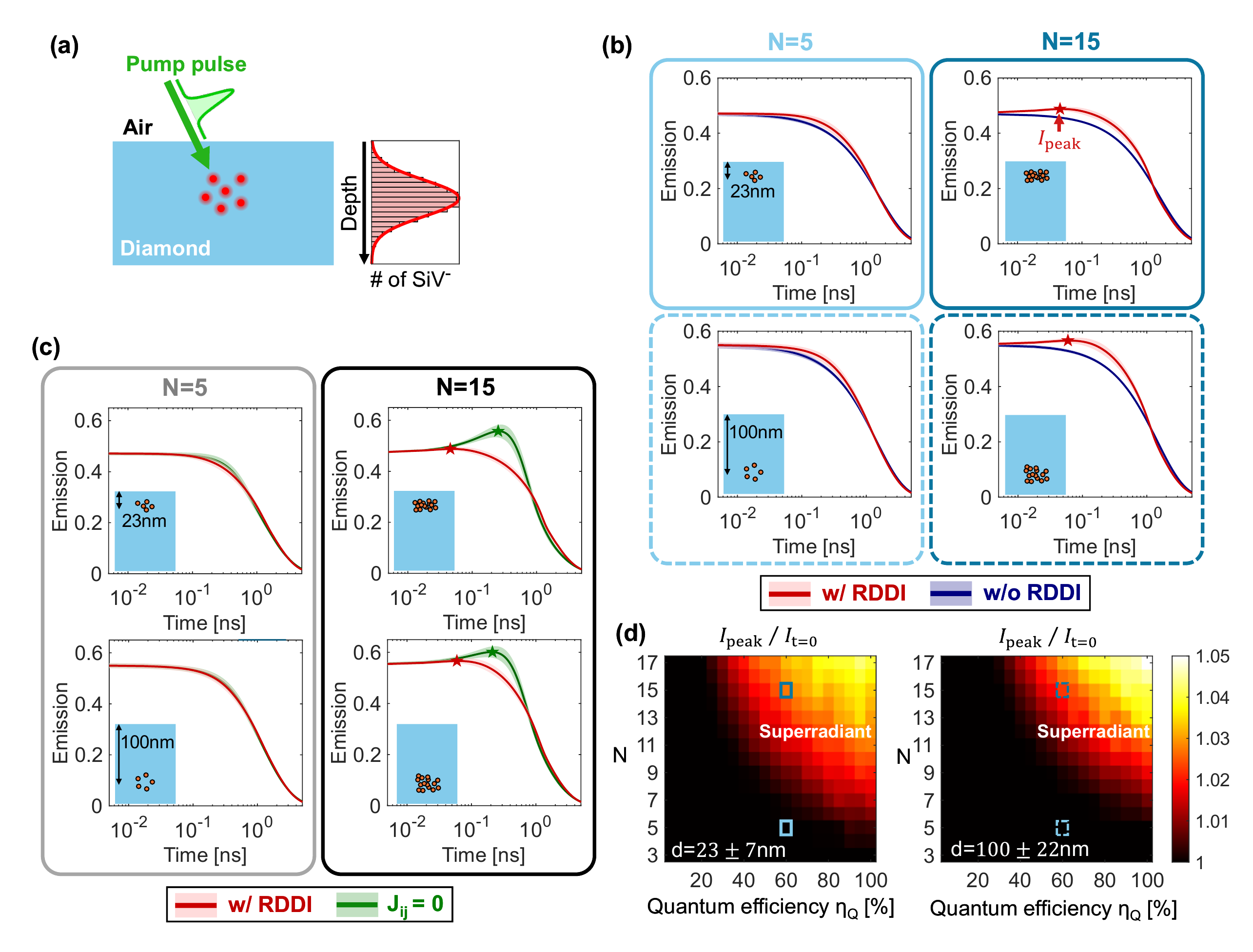}
  \caption{Time-resolved collective emission dynamics in SiV$^{-}$ ensembles. 
  (a) Schematic illustration of the TRPL experiment. A pulsed laser is used to excite the SiV$^{-}$ ensemble, and the time-evolution of the collected fluorescence signal is analyzed. In our simulations the depths of SiV$^{-}$ centers are sampled from a Gaussian distribution. 
  (b) Time-evolution of the photon emission rate, averaged over $300$ clusters. Shaded areas denote $\pm1$ standard deviation. Here ``w/o RDDI'' means that in Equation~(\ref{eq:eff_Hamiltonian})(\ref{eq:Lindblad}), both $J_{ij}$ and $\Gamma_{ij}$ are set to zero for $i\neq j$ pairs. 
  For $N=5$, the curve is almost indistinguishable from the case without dipole-dipole interaction, while for $N=15$ a superradiant burst emerges. 
  (c) Time-evolution of the photon emission rate, with the coherent part of interaction disabled ($J_{ij}=0$ in Equation~\ref{eq:eff_Hamiltonian}), while the dissipative part remains $\Gamma_{ij}\neq 0$. 
  Results are shown in green. For $N=15$, an apparent superradiant burst appears, with $I_\text{peak}/I_{t=0}\approx 1.173$ for $d=23\pm 7$ nm ($1.086$ for $d=100\pm 22$ nm). Removing $J_{ij}$ eliminates the dephasing induced by strong near-field interaction, thus strengthens the superradiance phenomenon. 
  (d) Maximum photon emission rate $I_\text{peak}$, normalized by $I_{t=0}$, across different parameter settings. The emitter number $N$ and quantum efficiency $\eta_{Q}$ are varied. Threshold behavior is observed for both the ``shallow'' and ``deep'' cases: sufficiently large $N$ and $\eta_{Q}$ are required to observe a superradiant peak. The four cases shown in panel (b) are marked by squares. }
  \label{fig:fig3}
\end{figure}
In this section, we use the developed cumulant expansion simulation framework to study the time-evolution of a fully inverted SiV$^{-}$ cluster, and to determine under what conditions a superradiant burst can emerge in the presence of realistic disorder. 
As predicted by R. Dicke in 1954, for a small cluster of totally inverted quantum emitters, the photon-mediated interactions make them synchronize and emit light coherently \cite{Dicke1954, Gross1982}. This will result in a cooperative acceleration of the decay process, leading to a superradiant burst of photon emission rate at time $t>0$. 
In the idealized Dicke model, the emitters are assumed to be indistinguishable from each other. Therefore, only the states that obey permutation symmetry need to be considered, which significantly reduces the computational cost. 
In contrast, a cluster of color centers fabricated through ion implantation is intrinsically disordered, and thus cannot be regarded as indistinguishable. As we have discussed, the nonidealities include: (1) the random distribution of emitter positions, which leads to dipole-dipole interactions between pairs of emitters, making the Hamiltonian asymmetric; (2) dipole moment orientations; (3) inhomogeneity, causing different emitters to have different resonance frequencies as well as decay rates. 
For a random cluster, these factors break the permutation symmetry, making it unclear when would a strong superradiance phenomenon emerge. 
Furthermore, prior research works have been unable to provide detailed understanding about the effect of cluster depth, cluster density, as well as quantum efficiency. 

In this part, we use the developed simulation framework to answer the questions posed above by simulating the time-evolution of SiV$^{-}$ clusters, in which all SiV$^{-}$ centers are initialized to occupy the excited state. 
This scenario closely resembles the setup of time-resolved photoluminescence (TRPL) experiments, where a pump laser pulse creates the population inversion and the subsequent fluorescence signal is recorded, as schematically shown in Figure 3a. 
The initial condition for solving the ODEs is as follows: at $t=0$, $\langle \hat{\sigma}_{i}^{22} \rangle = 1$ for all $i$'s, $\langle \hat{\sigma}_{i}^{22} \hat{\sigma}_{j}^{22} \rangle = 1$ for all $(i, j)$ pairs, while the expectation values of all other operators start from zero. Therefore all coherences vanish at $t=0$ \cite{Bigorda2023}. 
The decay process of the cluster is characterized by the population of the excited state, defined as $P_\text{exc}(t) = \sum_{i} \langle \hat{\sigma}_{i}^{22} \rangle (t)$. Based on $P_\text{exc}(t)$, the photon emission rate can be calculated as
\begin{equation}
    I(t) = -\frac{d P_\text{exc}(t)}{dt} - \sum_{i} \Gamma^\text{nrad}_{i} \langle \hat{\sigma}_{i}^{22} \rangle (t), 
\end{equation}
where the first term represents the total population decay, and the second term subtracts the contribution from non-radiative decay so that $I(t)$ reflects only radiative emission \cite{Novotny2006}. In the absence of non-radiative decay, this expression reduces to the ideal form $-\frac{d P_\text{exc}(t)}{dt}$ \cite{Bigorda2023, Damanet2016}. 
Figure 3b shows the simulated photon emission rate $I(t)$ obtained for clusters, with an intrinsic quantum efficiency of $\eta_{Q}=\frac{\Gamma^\text{rad}}{\Gamma^\text{rad} + \Gamma^\text{nrad}} \approx 60\%$ \cite{Moller2014, Bezard2024}, corresponding to non-radiative decay rates $\Gamma^\text{nrad}/2\pi = 46~\text{MHz}$. 
The left two panels present results for $N=5$ SiV$^{-}$ centers, while the right two panels for $N=15$. 
The top and bottom rows correspond to ``shallow'' clusters with depth $d=23\pm 7$ nm and ``deep'' clusters with $d=100\pm 22$ nm, respectively. The mean and standard deviation of these depth distributions are obtained through SRIM simulations (details are provided in Supplementary Note S4). 
For each of these 4 configurations we perform 300 Monte Carlo realizations in order to extract the ensemble averages (shown by the dark solid lines) as well as standard deviations (shown by the shaded areas). A convergence analysis of our Monte Carlo simulation has been provided in Supplementary Note S6. 
The photon emission rates plotted have been normalized by $N \Gamma^\text{rad}$. 
A superradiant burst can be recognized when the maximum photon emission rate $I_\text{peak}$ is larger than the emission rate at $t=0$, denoted as $I_{t=0}$. In the figures, $I_\text{peak}$ has been marked with an asterisk when it occurs at $t>0$. 
The data show that clusters with $N=15$ exhibit a superradiant burst under high quantum efficiency $\eta_{Q}=60\%$, while $N=5$ clusters do not. 
The shallow clusters show a reduced emission rate, which can be attributed to the modified local density of states near the air-diamond interface. Such depth dependence of the radiative decay rate is consistent with prior calculations of interface-induced lifetime variations \cite{Zahedian2023}. 
In order to demonstrate the effect of dipole-dipole interaction, we have repeated the same set of simulation, but turning off all interaction terms. Specifically, this is done by setting $J_{ij} = \Gamma_{ij} = 0$ for all $i\neq j$ in Equation~(\ref{eq:eff_Hamiltonian})(\ref{eq:Lindblad}). The corresponding results are displayed with blue curves. 
In this case without the dipole-dipole interaction, the superradiant burst is absent, since all SiV$^{-}$ centers emit photons independently. 

Although a superradiant burst emerges for $N=15$, its amplitude is small: both depth distributions shown in Figure 3b gives $I_\text{peak}/I_{t=0} < 1.03$. 
This is in sharp contrast to the ideal Dicke model, where the maximum emission rate is $I_\text{peak}/I_{t=0}=\frac{N}{4}+\frac{1}{2} = 4.25$ and scales as $\sim O(N)$ when $N$ increases \cite{Dicke1954}. 
Such suppressed superradiance arises from not only inhomogeneity, but also strong near-field RDDI in small clusters, as has been pointed out in prior theoretical works \cite{Friedberg1972, Friedberg1974, Gross1982, Venkatesh2018}. 
The coherent coupling term $J_{ij}\propto r_{ij}^{-3}$ becomes strong at short separation, leading to a strong dephasing mechanism that inhibits synchronization. Such suppression of superradiance has already been observed in numerical simulations involving interacting atoms \cite{Damanet2016}. 
A similar effect has been observed to limit the performance of quantum sensing applications: high dopant concentration leads to enhanced magnetic dipole–dipole interactions, thus reducing the achievable coherence time \cite{Zhou2020, Merkel2021, Gao2025}. 
Note that our proposed simulation framework is general enough to model this scenario as well. 
To isolate this dephasing effect, we set the coherent part of RDDI as $J_{ij}=0$ while retaining the cooperative decay $\Gamma_{ij}\neq 0$ in Equation~(\ref{eq:Lindblad}), then repeat the Monte Carlo simulations with other parameters unchanged. 
The results are shown in Figure 3c. 
For $N=5$ the results are nearly unchanged, whereas for $N=15$ both depth distributions show a more prominent superradiant burst, with $I_\text{peak}/I_{t=0}\approx 1.173$ for $d=23\pm 7$ nm and $1.086$ for $d=100\pm 22$ nm. 
In Supplementary Note S7 we have included a similar simulation with higher density (the depth distribution has a standard deviation of $1$ nm, which can be achieved through delta-doping), to further demonstrate the effect of near-field interaction. 
These simulations are consistent with theoretical predictions \cite{Damanet2016} and help explain why superradiance is difficult to observe in high-density clusters of emitters. Note that by engineering the photonic environment it is possible to mitigate this effect and achieve stronger superradiance \cite{Angerer2018, Pallmann2024, Moeller2014}. 

Based on the above observations, we next identify the conditions under which superradiance occurs. 
So far we have fixed the intrinsic quantum efficiency at $\eta_{Q}=60\%$. Reported values for the SiV$^{-}$ quantum efficiency, however, vary widely across the literature: early measurements suggested very low quantum efficiency ($\eta_{Q}\approx 5\%\sim 10\%$) \cite{Neu2012, Ruf2021, Bradac2019}, whereas more recent studies have reported $\eta_{Q}\approx 50\%\sim 60\%$ \cite{Moller2014, Bezard2024}. 
To quantify how the intrinsic quantum efficiency impacts collective emission, we further conduct a systematic parameter sweep, by varying $N$ (from $3$ to $17$) and quantum efficiency $\eta_{Q}$ ($5\%\sim 100\%$), for both ``shallow'' ($d=23\pm 7$ nm) and ``deep'' ($d=100\pm 22$ nm) clusters. 
Figure 3d presents a heatmap of the ratio between maximum emission rate $I_\text{peak}$ and the initial emission rate $I_{t=0}$. 
Each data point in the heatmap corresponds to the averaged result obtained from 120 cluster realizations. 
A clear boundary emerges: superradiant bursts occur only for sufficiently large number $N$ and high $\eta_{Q}$. 
This thresholded behavior remains qualitatively the same for both depth distributions. 
The cooperative enhancement of this superradiant burst is modest: even under the most favorable settings ($N=17$, $\eta_{Q}=100\%$) the relative difference between $I_\text{peak}$ and $I_{t=0}$ is only around $5\%$ . 
Again, this is much smaller compared to the ideal Dicke superradiance. 

In conclusion, our simulations demonstrate that small SiV$^{-}$ cluster can exhibit superradiant effect driven by RDDI, even in the presence of spatial disorder and inhomogeneous broadening. 
We also observe the suppression of superradiance caused by strong near-field interactions, which is consistent with theoretical predictions made in previous works. 
With Monte Carlo simulations, we identify two important factors that govern the amplitude of superradiant burst: the number of emitters $N$ and the intrinsic quantum efficiency $\eta_{Q}$. 
The above cumulant expansion framework therefore offers a versatile tool for predicting and understanding the collective phenomena in color center ensembles. 

\subsection{Photoluminescence excitation results}

\begin{figure}
  \centering
  \includegraphics[width=0.8\linewidth]{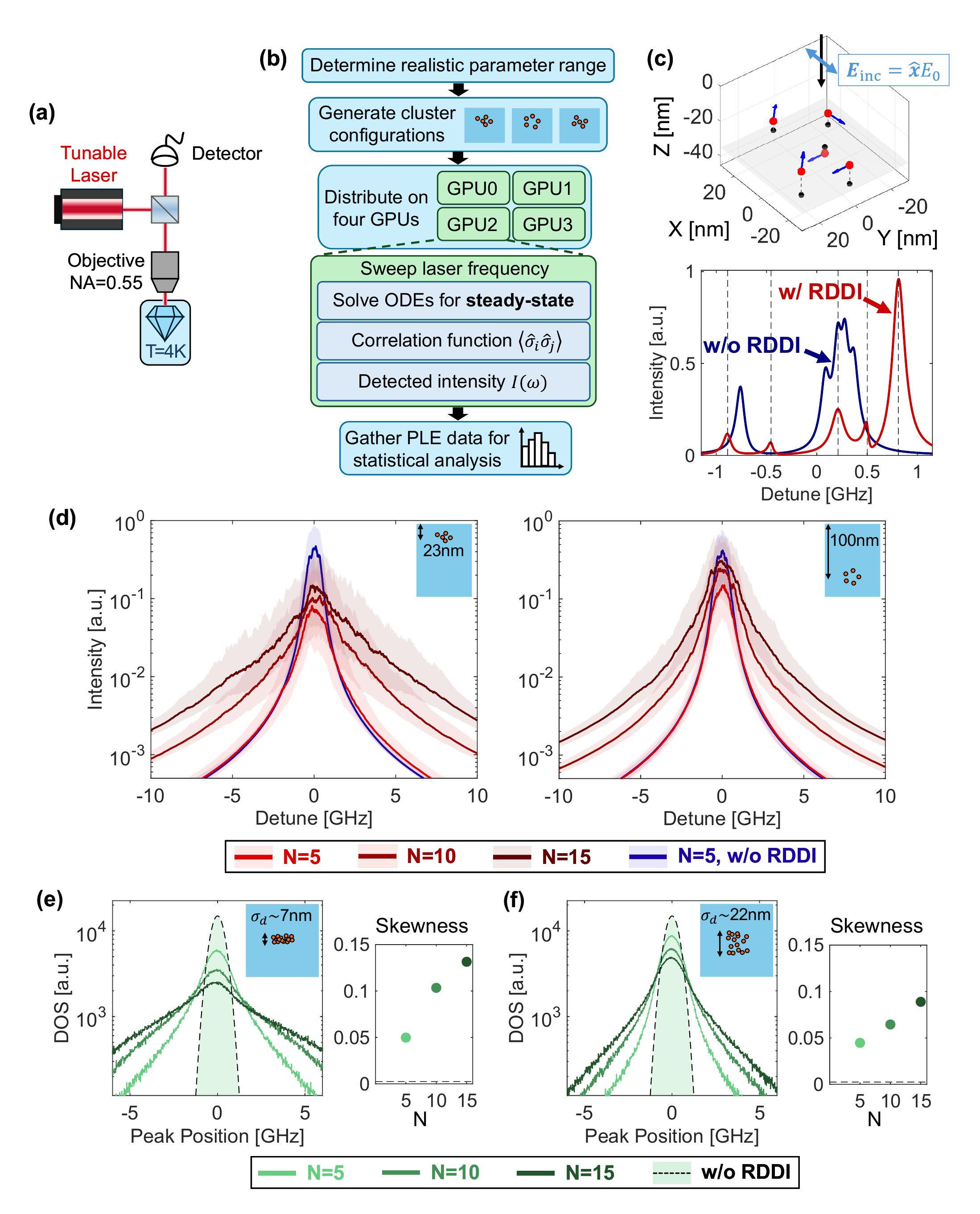}
  \caption{Collective behavior obtained from the frequency-domain response of SiV$^{-}$ ensembles. All simulations are assuming an intrinsic quantum efficiency of $\eta_{Q}=60\%$. 
  (a) Simplified schematic of the experimental setup used to measure the PLE spectrum. A tunable laser is used for resonant excitation. A single objective lens is used to focus the incident beam onto the sample, as well as to collect the emission. 
  (b) Flowchart of the simulation framework. We randomly generate $600$ cluster configurations. For each cluster the laser frequency is swept by varying the detune $\Delta$ and calculate the PLE spectrum. Finally, PLE spectra from all clusters are aggregated and analyzed statistically. 
  (c) An example with 5 color centers. The upper panel shows the positions (red dots) and the dipole moments (blue arrows) of all SiV$^{-}$ centers. The unit for $x$, $y$ and $z$ coordinates are all nanometer. The lower panel shows the PLE spectrum (blue) and compares it with the case without dipole-dipole interaction (red). 
  (d) The PLE spectra obtained from Monte Carlo simulation of $600$ clusters. The left panel corresponds to the ``shallow'' distribution ($23\pm 7$ nm), while the right panel corresponds to the ``deep'' distribution ($100\pm 22$ nm). Solid lines show the median; shaded bands indicate the $25\%$ to $75\%$ interquartile range. For comparison, blue curves show results when disabling RDDI. 
  (e)(f) Density of single-excitation states in a homogeneous environment. The standard deviation of depth distribution is (e) $\sigma_{d}=7$ nm and (f) $\sigma_{d}=22$ nm, respectively. 
  Solid lines show results obtained for $5, 10, 15$ emitters. 
  The shaded area indicates an ideal Gaussian distribution with a standard deviation of 400 MHz, corresponding to the case without dipole-dipole interactions. 
  The inset shows the skewness of the three distributions. As a reference, the dashed line marks the skewness of the ideal Gaussian, which is close to zero. }
  \label{fig:fig4}
\end{figure}
In this section, we aim to use our simulation framework to calculate the frequency-domain response of SiV$^{-}$ cluster.
Predicting spectral properties via numerical simulation is important, because there are many widely-used experimental techniques that do not resolve transient dynamics, but instead rely on measuring the steady-state emission under CW excitation. 
One important example is the PLE spectrum. The simplified schematic of a typical PLE measurement setup is shown in Figure 4a: a tunable narrow-band laser serves as the light source which excites the SiV$^{-}$ centers resonantly. 
The resulting fluorescence signal is then collected by an objective lens before entering a photodetector. 
By sweeping the laser frequency $\omega_{d}$ and recording the intensity of the collected signal $I(\omega_{d})$, one obtains the PLE spectrum, from which the positions of resonance peaks can be extracted directly. 
Although PLE spectroscopy is widely used for characterizing quantum emitters, the role of photon-mediated dipole-dipole interactions in shaping these spectra has received little attention. 
Here, we employ our cumulant expansion approach to quantify how dipole-dipole interaction modifies the PLE spectrum of disordered color center clusters. Thus we are able to identify observable difference that can occur in such measurements. 

The entire simulation workflow is summarized in the flow chart, as shown in Figure 4b. 
In order to overcome the sample-to-sample variation discussed at the beginning of this section, we have to generate hundreds of different cluster realizations, then perform an averaging through Monte Carlo simulations. This should enable direct comparison with experimental results in a statistical manner. 
Specifically, we examine three different number of emitters $N\in \{5, 10, 15\}$, as well as two depth distributions (namely the ``shallow'' case with $d=23\pm7$ nm, and the ``deep'' case with $d=100\pm22$ nm), similar to the previous part. 
Other parameters are kept exactly identical to those used in the time-domain analysis, if not mentioned. For each of these six scenarios, 600 random clusters are created. 
For every cluster configuration, we need to sweep the laser detuning $\Delta = \omega_{0} - \omega_{d}$. For each $\Delta$ the equations of motion are solved until reaching steady-state. 
This procedure leads to a very large number of ODE solving required in order to obtain faithful averaged results. 
To make the Monte Carlo simulations less time-consuming, we have implemented the second-order cumulant expansion approach using CUDA C, making it possible to accelerate the ODE solving part by around $30\sim 40\times$ when run on a GPU (more details about the implementation and the performance can be found in Supplementary Note S8). 
Finally, as shown in Figure 4b, the 600 generated SiV$^{-}$ cluster realizations are distributed across 4 NVIDIA Titan Xp graphic cards (each having 12GB memory), making the calculation of PLE spectrum computationally feasible. 

To obtain the detected PLE signal for a given laser detune $\Delta$, we first have to derive the relationship between the detected intensity and the expectation values of operators. 
We begin by expressing the electric field operator at the detector's position $\bm{r}_\text{det}$ via the dyadic Green's function \cite{Richter1990, Garcia2017a, Garcia2017b}:
\begin{equation}
    \hat{\bm{E} }^{+}(\bm{r}_\text{det}, \omega) = \mu_{0} \omega_{0}^{2} \sum_{i} \left( \hat{\sigma}_{i}^{12} \cdot \mathbf{G}(\bm{r}_\text{det}, \bm{r}_{i}) \cdot \bm{\mu}_{i} \right),
\end{equation}
so that the detected emission spectrum can be derived as \cite{Richter1990}
\begin{align}
    I(\bm{r}_\text{det}, \omega) &\triangleq \frac{2}{\eta_{0}} \left\langle \hat{\bm{E} }^{-}(\bm{r}_\text{det}, \omega) \cdot \hat{\bm{E} }^{+}(\bm{r}_\text{det}, \omega) \right\rangle \nonumber \\
    & = \frac{2\mu_{0}^{2} \omega_{0}^{4}}{\eta_{0}} \sum_{i, j} 
    \left( \mathbf{G^{}}(\bm{r}_\text{det}, \bm{r}_{i}) \cdot \bm{\mu}^{}_{i} \right)^{*} \cdot 
    \left( \mathbf{G}(\bm{r}_\text{det}, \bm{r}_{j}) \cdot \bm{\mu}_{j} \right)
    C_{ij}(\omega), 
\end{align}
where the correlation function $C_{ij}(\omega) = \int_{-\infty}^{\infty} d\tau 
\left\langle \hat{\sigma}_{i}^{21}(t) \hat{\sigma}_{j}^{12}(t+\tau) \right\rangle e^{i\omega \tau}$ has been introduced. 
This correlation function is computed by utilizing the quantum regression theorem, which requires us to first time-evolve the equation of motion with laser driving term $\Omega_{i}$ included, until the expectation value of all operators have reached steady state $\frac{d\langle \hat{O} \rangle}{dt} = 0$. The detailed procedure has been provided in Supplementary Note S9. 
Finally, the total detected intensity for a given $\Delta$ is obtained by integrating $I(\bm{r}_\text{det}, \omega)$ over the frequency range: 
\begin{equation}
    I_\text{det}(\Delta) = \int_{0}^{\sin^{-1}(\text{NA})} \sin\theta d\theta \int_{0}^{2\pi} d\phi \left[ \int d\omega     I(\bm{r}_\text{det}, \omega) \right].    
\end{equation}
Here we have assumed that the detector's responsivity does not depend on frequency, thus can be safely ignored from the integral. An integration over the solid angle has been included to take into account the finite numerical aperture (NA) of the objective lens. 
Notice that in all our simulations we take $\text{NA}=0.55$, which is consistent with the $50\times$ Mitutoyo Plan Apo objective \cite{Mitutoyo}. After sweeping over detune $\Delta$, the resulting $I_\text{det}(\Delta)$ gives a theoretical prediction of the PLE spectrum, which can be compared with experimental results. 

In order to obtain a qualitative understanding of the features shown in the PLE spectrum, Figure 4c presents a representative example with $N=5$ color centers. 
The upper panel displays the 3D positions of these SiV$^{-}$ centers (marked with red dots), together with their dipole moment orientations (marked with blue arrows). All $x$, $y$ and $z$ coordinates are shown in nanometers. 
The lower panel compares the calculated PLE spectra with (the red curve) and without (the blue curve) dipole-dipole interaction. 
Without dipole-dipole interaction, each color center emits independently. Therefore the calculated PLE spectrum corresponds to a summation of individual Lorentzian peaks contributed by each emitter. The resonance peaks remain close to each other, within the range determined by inhomogeneous broadening.  
On the contrary, when incorporating dipole-dipole interaction, the peak positions shown in the PLE spectrum are pushed further away from each other. We have marked the eigenfrequencies obtained from single-excitation eigenstates using dashed vertical lines (derivation details can be found in Supplementary Note S10). This verifies the correctness of our implementation, in the sense that under weak-excitation limit the resonance peaks shown in the PLE spectrum should coincide with the single-excitation eigenfrequencies. 

To capture the effects of disorder, again we perform Monte Carlo simulations. 
We compute PLE spectra for 600 cluster configurations, and the results are plotted in Figure 4d: the solid lines denote the median of PLE spectrum, and the shaded bands indicate the $25\%\sim75\%$ interquartile range. 
The two figure panels, correspond to ``shallow'' clusters ($d=23\pm 7$ nm, left) and ``deep'' clusters ($d=100\pm 22$ nm, right), respectively. 
Without dipole-dipole interactions, the independent-emitter case (as shown by the blue curves) reveals a Gaussian distribution of resonance peak positions, with a $400$ MHz standard deviation, reflecting the underlying inhomogeneous broadening. 
On the other hand, with dipole-dipole interactions included (as shown by the red curves), the median PLE spectra broaden substantially. Increasing cluster density (by increasing $N$) leads to stronger broadening. 
This arises because strong interactions $J_{ij}$ push resonance peaks away from the central frequency $\omega_{0}$. 
To attribute the broadening observed in Figure 4d, we next compute the single-excitation density of states (DOS), since PLE peak positions coincide with the single-excitation eigenfrequencies under weak-excitation. 
Importantly, we calculate the DOS inside a homogeneous environment with $n=2.4$ (derivation details can be found in Supplementary Note S10). The interface is not considered in order to isolate the intrinsic contribution of near-field dipole–dipole interactions, as well as to reduce computational cost. 
Results are shown in Figure 4e, 4f, with depth standard deviations $\sigma_{d}=7$ nm and $\sigma_{d}=22$ nm, respectively. 
Each histogram contains data from $7.5\times 10^{5}$ peaks. As before, three different emitter numbers, $N=5$, $N=10$ and $N=15$, are considered. 
As a reference, the light-green shaded area denotes an ideal Gaussian distribution with 400 MHz standard deviation (matching the inhomogeneous broadening of $\delta_{i}$), corresponding to the case without dipole-dipole interactions. 
As cluster density increases, which can be achieved by increasing the number $N$ or by imposing a narrower depth spread, the interaction-induced broadening becomes more pronounced compared to the independent-emitter case. 
Moreover, it has been predicted that sufficiently strong dipole-dipole interactions yield an asymmetric DOS profile \cite{Scholak2014}. We therefore evaluate the skewness $\mathbb{E}\left[\left(\frac{X-\mu}{\sigma}\right)^{3} \right]$ across all different scenarios (details can be found in Supplementary Note S11). 
Plotting the skewness values (dots in the insets) and comparing it with that of an ideal Gaussian distribution (dashed line) confirms that dipole-dipole interactions indeed lead to a biased peak distribution. This observation is consistent with previous work, in the sense that an asymmetric DOS profile can be predicted for an interacting Rydberg atom gases \cite{Scholak2014}. For denser clusters, the skewness increases due to stronger RDDI. A more systematic understanding of the DOS may be achieved via random-matrix theory \cite{Beenakker1997}, which we view as a promising direction for future analysis.

Experimentally, determining the interaction strength within a given cluster remains challenging (though feasible via multi-dimensional spectroscopy \cite{Day2022, Smallwood2021, Mukamel1995, Groll2025}). 
Nevertheless we believe that the simulated signatures, including broadening as well as non-zero skewness, should be observable in experiments. 
With current technologies, a practical route is to fabricate hundreds of clusters with different densities, then analyze the PLE peak distributions statistically. 
Our simulation framework should enable a direct comparison between measured data and statistical predictions. This will allow quantitative assessment of interaction strengths in SiV$^{-}$ ensembles, and could help interpret the effect of such interactions. 

\section{Conclusion \& Discussion}
In conclusion, in this work we have performed detailed modeling and analysis of an ensemble of quantum emitters, especially focusing on the collective behavior induced by photon-mediated dipole-dipole interactions. By extending the cumulant expansion approach, we have established a simulation framework, reducing the computational cost down to $O(N^3)$ for a given emitter number $N$. 
Without loss of generality, we then focus on SiV$^{-}$ color centers. The framework incorporates dipole–dipole interaction from a structured photonic environment as well as key experimental nonidealities, including inhomogeneous broadening, spatial disorder, and random orientations. 
The above factors have made it difficult for the community to go beyond single-emitter regime and understand the effect of RDDI. 
In order to overcome this issue, we have combined our numerical simulation technique with Monte Carlo sampling, by obtaining the averaged response over hundreds of SiV$^{-}$ cluster realizations, thanks to the highly efficient implementation. 
Since it is possible to obtain statistically averaged properties of a large number of clusters in experiment, we believe that this type of measured data can be compared directly with the results obtained from Monte Carlo simulations, making our approach a powerful predictive tool. 

More specifically, by simulating the dynamics of such emitter clusters, we have extracted two observable collective phenomena caused by dipole-dipole interaction. 
We first simulate the photon emission process of an excited cluster. 
Despite the disorder among different color centers, superradiant burst can appear once the number of color centers $N$ exceeds a certain threshold. 
Another key factor is the quantum efficiency, which should be high enough in order to observe a burst. 
We also observe that, beside inhomogeneous broadening, near-field interaction also suppresses superradiance, which is consistent with prior predictions. 
Secondly, by calculating the intensity of photon emission under CW laser excitation, the PLE spectrum can be predicted numerically. 
With the presence of dipole-dipole interaction, the distribution of peak positions becomes much wider. 
Such interaction-induced broadening grows with cluster density and can become distinguishable compared to static inhomogeneous broadening, thus serves as a robust fingerprint of RDDI. 
The DOS exhibits a non-zero skewness that increases with cluster density. 
This makes it possible to infer the presence and strength of RDDI experimentally, by measuring PLE peak distributions across clusters with different densities. 

The simulation framework is not restricted to modeling color centers: any ensemble of quantum emitters, including atoms, molecules, quantum dots, or other solid-state defects, can be treated with the same formalism, so long as the parameters have been chosen accordingly. 
Therefore, we anticipate that the collective phenomena identified here, including superradiant emission and interaction-induced broadening, will appear in a range of state-of-the-art platforms. We also believe that the framework can serve as a practical tool for analyzing experimental results when seeking to detect collective effects in disordered systems. 

\medskip
\textbf{Supporting Information} \par 
Supporting Information is available from the Wiley Online Library or from the author.

\medskip
\textbf{Acknowledgements} \par 
This material is based upon work supported by National Science Foundation Grant No. 2016136 for the QLCI center Hybrid Quantum Architectures and Networks. The authors would like to thank Prof. Shimon Kolkowitz, Jietian Liu, Ming Zhou, and Minjeong Kim for insightful discussions. 

\medskip
\textbf{Conflict of Interest} \par 
The authors declare no conflict of interest.

\medskip
\textbf{Data Availability Statement} \par 
All codes for this study and the resulting data will be made available in a public repository upon acceptance.

\medskip

%
\bibliography{refs}

\clearpage

\appendix
\section{Dyadic Green's functions in inhomogeneous environment}
In this part we first clarify the definition of the dyadic Green's function $\mathbf{G}(\bm{r}, \bm{r}')$ appeared in the main text. The equations used to calculate the Green's function are also provided. 
Consider an inhomogeneous photonic environment described by relative permittivity $\epsilon_{r}(\bm{r})$ and relative permeability $\mu_{r}(\bm{r})$. Working at angular frequency $\omega_{0}$, the Maxwell's equations in frequency domain can be simplified as 
\begin{equation}
    \nabla \times \nabla \times \bm{E} - k^{2}\bm{E} = i\omega_{0} \mu_{0} \bm{J}, 
\label{eq:maxwell}
\end{equation}
where $k=\sqrt{\epsilon_{r} \mu_{r}} \omega_{0} / c_{0}$ denotes the wave vector inside material, and $\bm{J}$ denotes the electric current density. Notice that $\exp(-i\omega t)$ time dependence has been chosen. 
The dyadic Green's function of the above equation is defined as \cite{Novotny2006}
\begin{equation}
    \nabla \times \nabla \times \bm{G}(\bm{r}, \bm{r}') - k^{2}\bm{G}(\bm{r}, \bm{r}') = \delta(\bm{r}-\bm{r}') \mathbbm{1},
\end{equation}
where $\mathbbm{1}$ denotes the $3\times3$ identity matrix. Thus, the solution of eq.~(\ref{eq:maxwell}) can be expressed using the following integral:
\begin{equation}
    \bm{E}(\bm{r}) = i\omega_{0} \mu_{0} \int d^{3}r' \bm{G}(\bm{r}, \bm{r}') \cdot \bm{J}(\bm{r}'). 
\label{eq:E_GJ}
\end{equation}
Starting from a homogeneous environment. The dyadic Green's function shows translational invariance: $\bm{G}(\bm{r}, \bm{r}') = \bm{G}(\bm{r}-\bm{r}', \bm{0})$. The dyadic Green's function can be solved analytically: 
\begin{equation}
    \bm{G}(\bm{r}-\bm{r}', \bm{0}) = \frac{\exp(ikr)}{4\pi r} \cdot \left[(3\hat{r}\hat{r}-\mathbbm{1})(\frac{1}{k^{2}r^{2}} + \frac{1}{ikr}) - (\hat{r}\hat{r}-\mathbbm{1}) \right], 
    \label{eq:analytic_G}
\end{equation}
where $r=|\bm{r}-\bm{r}'|$ denotes the distance between two points, and $\hat{r}=(\bm{r}-\bm{r}')/r$. 

Next, we proceed and take into account the effect of interface between two media. The derivations in this section follows Ref.~\citenum{Hanson2008}. Consider a 3-D space filled with two media. The upper half-space ($z>0$) is filled with medium 1 (relative permittivity $\epsilon_{1}$), while the lower half-space ($z<0$) is filled with medium 2 (relative permittivity $\epsilon_{2}$). The corresponding wave vectors inside media $i$ can be defined as $k_{i}^{2} = \epsilon_{i} k_{0}^{2}$ ($i=1,2$). 
Without loss of generality, we now assume that both the source point $\bm{r}'=(x', y', z')$ and the inspection position $\bm{r}=(x, y, z)$ are inside medium 2, satisfying $z<0$ and $z'<0$. In order to obtain an expression for dyadic Green's function, we utilize the Hertzian potential approach \cite{Chew1999}. Specifically, the $\bm{E}$ field can be calculated using the equation $\bm{E} = \nabla(\nabla \cdot \bm{\pi}) + k_{2}^{2} \bm{\pi}$, where the Hertzian vector potential $\bm{\pi}(\bm{r})$ can be calculated using the Hertzian potential Green's dyadic $\bm{g}(\bm{r}, \bm{r}')$: 
\begin{equation}
    \bm{\pi}(\bm{r}) = \int d^{3}r' \bm{g}(\bm{r}, \bm{r}') \cdot \frac{\bm{J}(\bm{r}')}{-i\omega_{0} \epsilon_{2} \epsilon_{0}}.
\end{equation}
Based on the above equations, together with eq.~(\ref{eq:E_GJ}), the relationship between dyadic Green's function $\bm{G}(\bm{r}, \bm{r}')$ and $\bm{g}(\bm{r}, \bm{r}')$ is
\begin{equation}
    G_{ij} = \frac{1}{k_{2}^{2}} \left[ \frac{\partial}{\partial x}(\frac{\partial g_{jx}}{\partial i}) + \frac{\partial}{\partial y}(\frac{\partial g_{jy}}{\partial i}) + \frac{\partial}{\partial z}(\frac{\partial g_{jz}}{\partial i}) \right] + g_{ij}, ~\text{where}~i,j=x,y,z. 
\end{equation}
In order to find the analytical form of $G_{ij}$, we pay attention to the fact that $\bm{g}$ can be divided into two parts: 
\begin{equation}
    \bm{g}(\bm{r}, \bm{r}') = \bm{g}_{p}(\bm{r}, \bm{r}') + \bm{g}_{s}(\bm{r}, \bm{r}'),
\end{equation}
where $\bm{g}_{p}(\bm{r}, \bm{r}')$ stands for the ``principal" part inside homogeneous environment, while $\bm{g}_{s}(\bm{r}, \bm{r}')$ stands for the ``scattered'' part caused by the reflection at the interface. Both $\bm{g}_{p}(\bm{r}, \bm{r}')$ and $\bm{g}_{s}(\bm{r}, \bm{r}')$ are $3\times 3$ tensors:
\begin{equation}
    \bm{g}_{p}(\bm{r}, \bm{r}') = \frac{\exp(ik_{2}|\bm{r}-\bm{r}'|)}{4\pi |\bm{r}-\bm{r}'|} \mathbbm{1},
\end{equation}
\begin{equation}
    \bm{g}_{s}(\bm{r}, \bm{r}') = \begin{bmatrix}
    g_{t} & 0 & \frac{\partial}{\partial x} g_{c}\\
    0 & g_{t} & \frac{\partial}{\partial y} g_{c}\\
    0 & 0 & g_{n}
    \end{bmatrix}. 
\end{equation}
Both $g_{t}(\bm{r}, \bm{r}')$, $g_{n}(\bm{r}, \bm{r}')$ and $g_{c}(\bm{r}, \bm{r}')$ can be calculated using Sommerfeld integral over the in-plane wave vector $q$. By defining $p_{i}^{2}=q^{2}-k_{i}^{2}$ ($i=1,2$) and the in-plane distance $\rho=\sqrt{(x-x')^{2} + (y-y')^{2}}$, these three scalar fields can be derived as:
\begin{equation}
    g_{t}(\bm{r}, \bm{r}') = \frac{1}{2\pi} \int_{0}^{+\infty} \frac{p_{1}-p_{2}}{p_{1}+p_{2}} e^{p_{2}(z+z')} \cdot \frac{J_{0}(q\rho)}{2p_{2}} \cdot q dq,
\end{equation}

\begin{equation}
    g_{c}(\bm{r}, \bm{r}') = \frac{1}{2\pi} \int_{0}^{+\infty} \frac{2(\frac{\epsilon_{1}}{\epsilon_{2}}-1)p_{2}}{(p_{1}+p_{2})(\frac{\epsilon_{1}}{\epsilon_{2}}p_{2}+p_{1})} e^{p_{2}(z+z')} \cdot \frac{J_{0}(q\rho)}{2p_{2}} \cdot q dq,
\end{equation}

\begin{equation}
    g_{n}(\bm{r}, \bm{r}') = \frac{1}{2\pi} \int_{0}^{+\infty} \frac{\frac{\epsilon_{1}}{\epsilon_{2}}p_{2}-p_{1}}{\frac{\epsilon_{1}}{\epsilon_{2}}p_{2}+p_{1}} e^{p_{2}(z+z')} \cdot \frac{J_{0}(q\rho)}{2p_{2}} \cdot q dq.
\end{equation}

To verify that our implementation is correct, we've calculated the radiative lifetime of a dipole source inside diamond. Specifically, we benchmark our implementation by calculating the lifetime $\tau(d)$, where $d$ stands for the depth. The lifetime is normalized by the lifetime $\tau_{0}$ inside homogeneous environment. The results are consistent with the data displayed in Figure 2b of Ref.~\citenum{Zahedian2023}. We observe that when the dipole moment is perpendicular to the interface ($\theta_{e} = 0^{\circ}$), the lifetime can become much longer than $\tau_{0}$ if the dipole is shallow. On the other hand, when the dipole moment is parallel to the interface ($\theta_{e} = 90^{\circ}$), the lifetime $\tau$ does not differ much from $\tau_{0}$. 
\begin{figure}[h]
\centering
\includegraphics[width=1\linewidth]{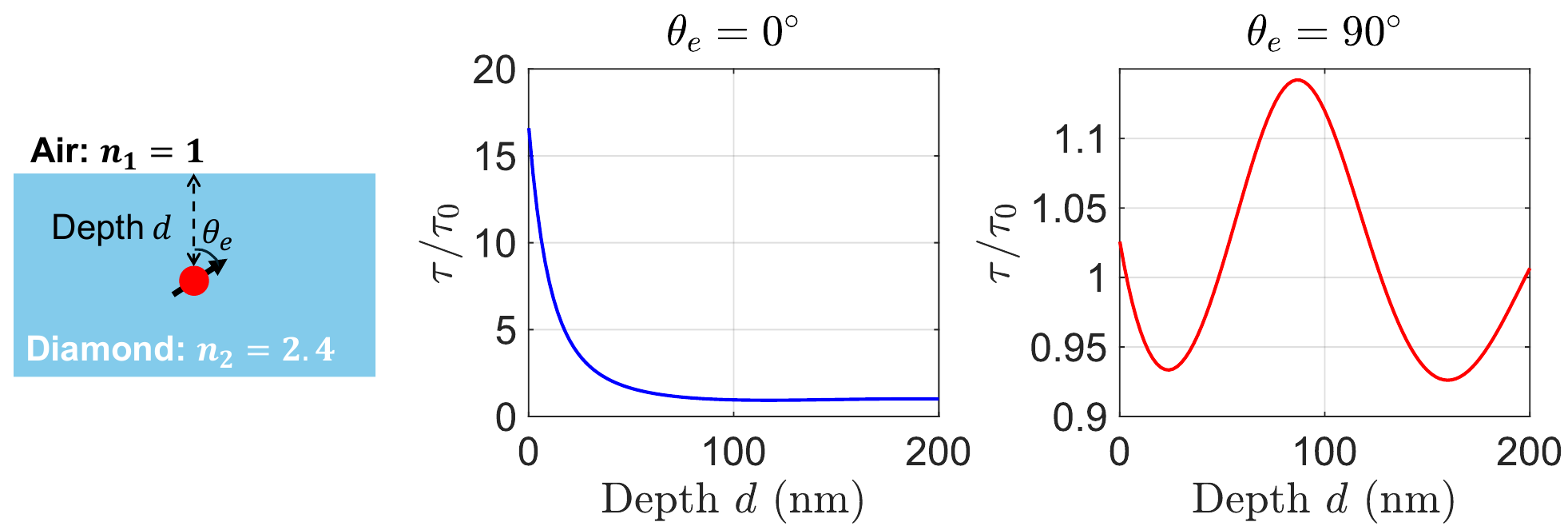}
\caption{Radiative lifetime $\tau$ of an electric dipole inside diamond, visualized as a funcntion of depth $d$. Two different cases are considered: the blue curve ($\theta_{e} = 0^{\circ}$) corresponds to a dipole that's perpendicular to the interface, while the red curve ($\theta_{e} = 90^{\circ}$) corresponds to a dipole that's parallel to the interface. The lifetime has been normalized using $\tau_{0}$, which stands for the lifetime inside homogeneous environment. The wavelength is chosen as $\lambda_{0} = 637$ nm, corresponding to the zero-phonon line of NV$^{-}$ center inside diamond.}
\label{fig:lifetime_ratio}
\end{figure}

~
\newpage
\section{Cumulant expansion: equations of motion}
Under the cumulant expansion framework, a closed set of coupled ordinary differential equations (ODEs) can be obtained. In the main text, we only present the ODEs related to the time-evolution of first-order operators $\hat{\sigma}^{12}_i$ and $\hat{\sigma}^{22}_i$. Here the equations related to second-order cumulants are presented:
\begin{align}
\frac{d}{dt}\langle \hat{\sigma}^{21}_i \hat{\sigma}^{12}_j \rangle = 
&\left[i(\delta_i - \delta_j) - \frac{\Gamma_{ii} + \Gamma_{i}^{\text{nrad}}}{2} - \frac{\Gamma_{jj} + \Gamma_{j}^{\text{nrad}}}{2}\right] \langle \hat{\sigma}^{21}_i \hat{\sigma}^{12}_j \rangle \nonumber \\
&+ \frac{\Gamma_{ji}}{2} \left[4 \langle \hat{\sigma}^{22}_i \hat{\sigma}^{22}_j \rangle - \langle \hat{\sigma}^{22}_i \rangle - \langle \hat{\sigma}^{22}_j \rangle \right]  +{iJ_{ji}( \hat{\sigma}^{22}_j - \hat{\sigma}^{22}_i ) } \nonumber \\
&+ \sum_{k \neq i,j} \left(iJ_{ki} - \frac{\Gamma_{ki}}{2}\right) \left[\langle \hat{\sigma}^{21}_k \hat{\sigma}^{12}_j \rangle - 2\langle \hat{\sigma}^{21}_k \hat{\sigma}^{22}_i \hat{\sigma}^{12}_j \rangle \right] \nonumber \\
&+ \sum_{k \neq i,j} \left(-iJ_{jk} - \frac{\Gamma_{jk}}{2}\right) \left[\langle \hat{\sigma}^{12}_k \hat{\sigma}^{21}_i \rangle - 2\langle \hat{\sigma}^{12}_k \hat{\sigma}^{21}_i \hat{\sigma}^{22}_j \rangle \right] \nonumber \\
&+ \frac{i\Omega_i^\ast}{2} \left(\langle \hat{\sigma}^{12}_j \rangle - 2\langle \hat{\sigma}^{22}_i \hat{\sigma}^{12}_j \rangle \right) - \frac{i\Omega_j}{2} \left(\langle \hat{\sigma}^{21}_i \rangle - 2\langle \hat{\sigma}^{21}_i \hat{\sigma}^{22}_j \rangle \right).
\end{align}
\begin{align} 
\frac{d}{dt} \langle \hat{\sigma}^{21}_i \hat{\sigma}^{22}_j \rangle &= \left[i(\Delta + \delta_i) - \frac{\Gamma_{ii} + \Gamma_{i}^{\text{nrad}}}{2} - (\Gamma_{jj} + \Gamma_{j}^{\text{nrad}})\right] \langle \hat{\sigma}^{21}_i \hat{\sigma}^{22}_j \rangle + \left[-iJ_{ji} - \frac{\Gamma_{ji}}{2}\right] \langle \hat{\sigma}^{22}_i \hat{\sigma}^{21}_j \rangle \notag \\
&+ \sum_{k \neq i,j} \left(iJ_{ki} - \frac{\Gamma_{ki}}{2}\right) \left[\langle \hat{\sigma}^{21}_k \hat{\sigma}^{22}_j \rangle - 2 \langle \hat{\sigma}^{21}_k \hat{\sigma}^{22}_i \hat{\sigma}^{22}_j \rangle \right] \notag  \\
&+ \sum_{k \neq i,j} \left(iJ_{kj} - \frac{\Gamma_{kj}}{2}\right) \langle \hat{\sigma}^{21}_k \hat{\sigma}^{21}_i \hat{\sigma}^{12}_j \rangle + \sum_{k \neq i,j} \left(-iJ_{jk} - \frac{\Gamma_{jk}}{2}\right) \langle \hat{\sigma}^{12}_k \hat{\sigma}^{21}_i \hat{\sigma}^{21}_j \rangle \notag  \\
&+ \frac{i\Omega_i^\ast}{2} \left(\langle \hat{\sigma}^{22}_j \rangle - 2 \langle \hat{\sigma}^{22}_i \hat{\sigma}^{22}_j \rangle \right) + \frac{i\Omega_j^\ast}{2} \langle \hat{\sigma}^{21}_i \hat{\sigma}^{12}_j \rangle - \frac{i\Omega_j}{2} \langle \hat{\sigma}^{21}_i \hat{\sigma}^{21}_j \rangle.
\end{align}
\begin{align}
\frac{d}{dt} \langle \hat{\sigma}^{12}_i \hat{\sigma}^{12}_j \rangle = 
&\left[-i(2\Delta + \delta_i + \delta_j) - \frac{\Gamma_{ii} + \Gamma_{i}^{\text{nrad}}}{2} - \frac{\Gamma_{jj} + \Gamma_{j}^{\text{nrad}}}{2}\right] \langle \hat{\sigma}^{12}_i \hat{\sigma}^{12}_j \rangle \nonumber \\
&+ \sum_{k \neq i,j} \left(iJ_{ik} + \frac{\Gamma_{ik}}{2}\right) 
\left[2\langle \hat{\sigma}^{12}_k \hat{\sigma}^{22}_i \hat{\sigma}^{12}_j \rangle - \langle \hat{\sigma}^{12}_k \hat{\sigma}^{12}_j \rangle \right] \nonumber \\
&+ \sum_{k \neq i,j} \left(iJ_{jk} + \frac{\Gamma_{jk}}{2}\right) 
\left[2\langle \hat{\sigma}^{12}_k \hat{\sigma}^{12}_i \hat{\sigma}^{22}_j \rangle - \langle \hat{\sigma}^{12}_k \hat{\sigma}^{12}_i \rangle \right] \nonumber \\
&+ \frac{i\Omega_i}{2} \left(2\langle \hat{\sigma}^{22}_i \hat{\sigma}^{12}_j \rangle - \langle \hat{\sigma}^{12}_j \rangle \right) + \frac{i\Omega_j}{2} \left(2\langle \hat{\sigma}^{12}_i \hat{\sigma}^{22}_j \rangle - \langle \hat{\sigma}^{12}_i \rangle \right).
\end{align}
\begin{align}
\frac{d}{dt} \langle \hat{\sigma}^{12}_i \hat{\sigma}^{22}_j \rangle &= 
\left[-i(\Delta + \delta_i) - \frac{\Gamma_{ii} + \Gamma_{i}^{\text{nrad}}}{2} - (\Gamma_{jj} + \Gamma_{j}^{\text{nrad}})\right] \langle \hat{\sigma}^{12}_i \hat{\sigma}^{22}_j \rangle 
+ \left[iJ_{ij} - \frac{\Gamma_{ij}}{2}\right] \langle \hat{\sigma}^{22}_i \hat{\sigma}^{12}_j \rangle \nonumber \\
&\quad + \sum_{k \neq i,j} \left(iJ_{kj} - \frac{\Gamma_{kj}}{2}\right) \langle \hat{\sigma}^{21}_k \hat{\sigma}^{12}_i \hat{\sigma}^{12}_j \rangle 
+ \sum_{k \neq i,j} \left(-iJ_{jk} - \frac{\Gamma_{jk}}{2}\right) \langle \hat{\sigma}^{12}_k \hat{\sigma}^{12}_i \hat{\sigma}^{21}_j \rangle \nonumber \\
&\quad + \sum_{k \neq i,j} \left(iJ_{ik} + \frac{\Gamma_{ik}}{2}\right) 
\left[2\langle \hat{\sigma}^{12}_k \hat{\sigma}^{22}_i \hat{\sigma}^{22}_j \rangle - \langle \hat{\sigma}^{12}_k \hat{\sigma}^{22}_j \rangle \right] \nonumber \\
&\quad + \frac{i\Omega_i}{2} \left(2\langle \hat{\sigma}^{22}_i \hat{\sigma}^{22}_j \rangle - \langle \hat{\sigma}^{22}_j \rangle\right) 
+ \frac{i\Omega_j^\ast}{2} \langle \hat{\sigma}^{12}_i \hat{\sigma}^{12}_j \rangle 
- \frac{i\Omega_j}{2} \langle \hat{\sigma}^{12}_i \hat{\sigma}^{21}_j \rangle.
\end{align}
\begin{align}
\frac{d}{dt} \langle \hat{\sigma}^{22}_i \hat{\sigma}^{22}_j \rangle &= 
-(\Gamma_{ii} + \Gamma_{i}^{\text{nrad}} + \Gamma_{jj} + \Gamma_{j}^{\text{nrad}}) \langle \hat{\sigma}^{22}_i \hat{\sigma}^{22}_j \rangle \nonumber \\
&+ \frac{i\Omega_i^\ast}{2} \langle \hat{\sigma}^{12}_i \hat{\sigma}^{22}_j \rangle 
- \frac{i\Omega_i}{2} \langle \hat{\sigma}^{21}_i \hat{\sigma}^{22}_j \rangle 
+ \frac{i\Omega_j^\ast}{2} \langle \hat{\sigma}^{22}_i \hat{\sigma}^{12}_j \rangle 
- \frac{i\Omega_j}{2} \langle \hat{\sigma}^{22}_i \hat{\sigma}^{21}_j \rangle \notag \\
&+ \sum_{k \neq i,j} \left(iJ_{ki} - \frac{\Gamma_{ki}}{2}\right) \langle \hat{\sigma}^{21}_k \hat{\sigma}^{12}_i \hat{\sigma}^{22}_j \rangle + \sum_{k \neq i,j} \left(iJ_{kj} - \frac{\Gamma_{kj}}{2}\right) \langle \hat{\sigma}^{21}_k \hat{\sigma}^{22}_i \hat{\sigma}^{12}_j \rangle \notag \\
&+ \sum_{k \neq i,j} \left(-iJ_{ik} - \frac{\Gamma_{ik}}{2}\right) \langle \hat{\sigma}^{12}_k \hat{\sigma}^{21}_i \hat{\sigma}^{22}_j \rangle 
+ \sum_{k \neq i,j} \left(-iJ_{jk} - \frac{\Gamma_{jk}}{2}\right) \langle \hat{\sigma}^{12}_k \hat{\sigma}^{22}_i \hat{\sigma}^{21}_j \rangle.
\end{align} 
Note that in order to truncate the above equations, all expectations related to third-order operators shall be replaced by
\begin{equation}
    \langle \hat{O}_{1} \hat{O}_{2} \hat{O}_{3} \rangle 
\approx \langle \hat{O}_{1} \rangle \langle \hat{O}_{2} \hat{O}_{3} \rangle 
+ \langle \hat{O}_{2} \rangle \langle \hat{O}_{1} \hat{O}_{3} \rangle 
+ \langle \hat{O}_{3} \rangle \langle \hat{O}_{1} \hat{O}_{2} \rangle 
- 2 \langle \hat{O}_{1} \rangle \langle \hat{O}_{2} \rangle \langle \hat{O}_{3} \rangle,
\end{equation}
as explained in the main text. In our MATLAB implementation, the above ODEs are solved numerically with the help of MATLAB's ``ode45()'' function. 

\newpage
\section{Benchmarking the cumulant expansion implementation}
\begin{figure}[h]
  \centering
  \includegraphics[width=0.8\linewidth]{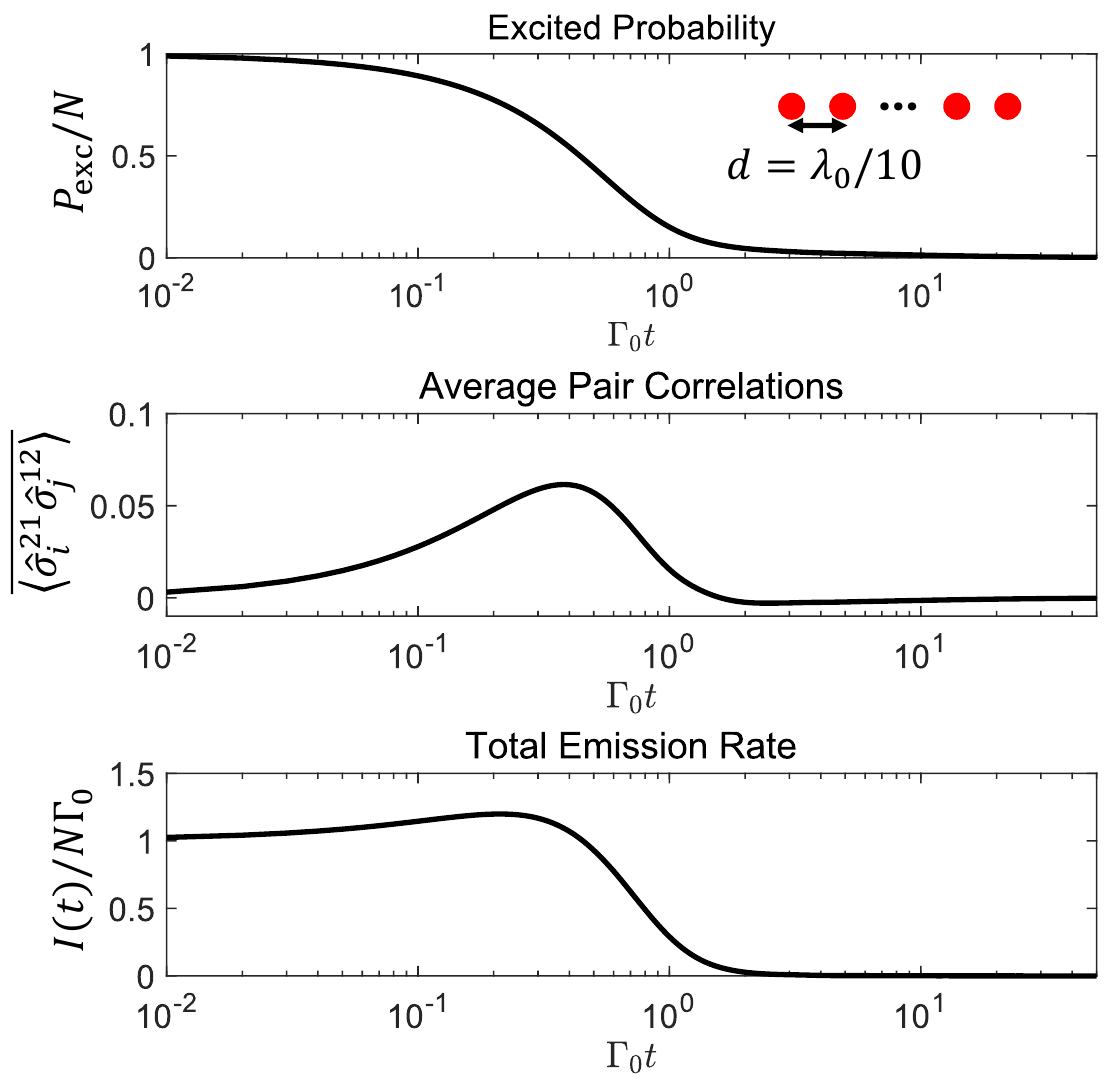}
  \caption{Time-evolution of an atom chain with $N=10$ atoms, obtained via cumulant expansion. (a) The excited state population $P_\text{exc}$, normalized by $N$. (b) The averaged pair correlations $\overline{ \langle \hat{\sigma}_{i}^{21} \hat{\sigma}_{j}^{12} \rangle }$. (c) The total photon emission rate $I(t)$, normalized by $N\Gamma_{0}$. }
  \label{fig:benchmark}
\end{figure}

In this part, the correctness of our implementation is verified by reproducing the example in Ref.~\citenum{Bigorda2023}. More specifically, we focus on the superradiance behavior of a fully-inverted atom chain which contains $N=10$ two-level systems (TLSs). The chain is aligned with the $x$ axis, while the dipole moment of all TLSs are pointing towards $z$ direction. All TLSs are identical, with the same radiative decay rate $\Gamma_{0}$. Similar to the superradiance examples given in the main text, the initial condition for solving the ODEs is set as follows: at $t=0$, $\langle \hat{\sigma}_{i}^{22} \rangle = 1$ for all $i$'s, $\langle \hat{\sigma}_{i}^{22} \hat{\sigma}_{j}^{22} \rangle = 1$ for all $(i, j)$ pairs, while the expectation values of all other operators start from zero. The simulation results obtained through cumulant expansion are displayed in Figure~\ref{fig:benchmark}. 
Three quantities have been visualized, namely:
the population of the excited state $P_{\text{exc}}(t)$, normalized by the number $N$; 
the averaged pair correlations $\overline{ \langle \hat{\sigma}_{i}^{21} \hat{\sigma}_{j}^{12} \rangle }$; 
the total emission rate $I(t) = -\frac{d P_\text{exc}(t)}{dt}$. The emission rate shows a superradiant peak at $t>0$, together with a buildup of the atomic coherences. Our simulation results match perfectly with figure 2 in Ref.~\citenum{Bigorda2023}, justifying the correctness of our implementation. 

\newpage
\section{Spatial distribution: SRIM simulation}
\begin{figure}[h]
  \centering
  \includegraphics[width=1\linewidth]{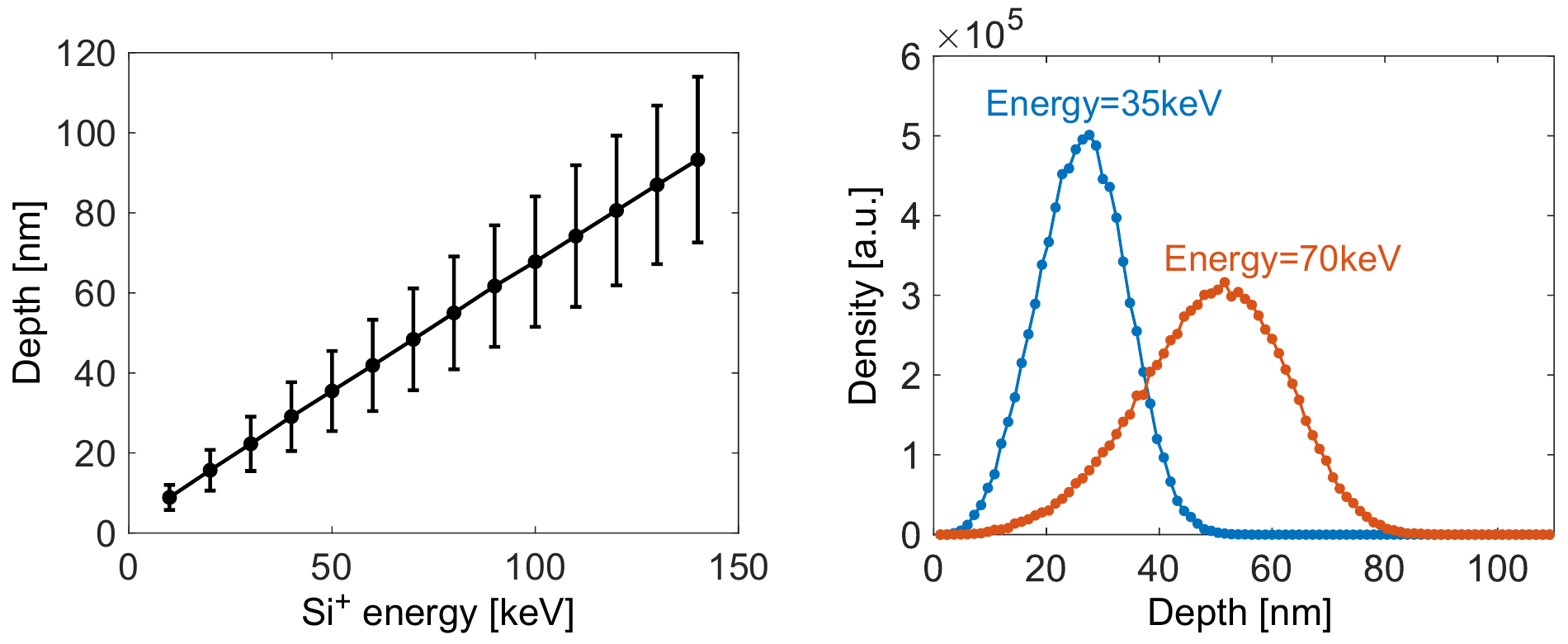}
  \caption{Depth distribution obtained through SRIM simulations. (a) The average depth versus incident Si$^{+}$ energy. The error bars indicate longitudinal straggle. (b) Density distributions at $35$ and $70~\mathrm{keV}$. }
  \label{fig:depth}
\end{figure}

In this part, we justify our choice of SiV$^{-}$ depth distribution. 
We compute the distribution by simulating Si implantation into diamond (density $3.51~\mathrm{g/cm^3}$) using SRIM~\cite{Ziegler2010}, sweeping incident Si$^{+}$ energies from $10$ to $140~\mathrm{keV}$ to map the mean depth and spread versus energy. 
To suppress channeling in diamond, the incidence is tilted by $7^\circ$. 
For each energy, we extract the depth histogram and report the mean depth as well as the straggle (i.e., the standard deviation along the $z$ direction). 
The left panel of Figure~\ref{fig:depth} shows the mean depth versus incident Si$^{+}$ energy with error bars as the longitudinal straggle; the right panel shows density distributions at $35$ and $70~\mathrm{keV}$, illustrating that higher energy yields a deeper and broader distribution. The depth profiles are well fitted by Gaussians, so we summarize them as $d=\text{mean}\pm\text{std}$. Based on these SRIM results, the main text adopts $d=23\pm 7~\mathrm{nm}$ for ``shallow'' clusters and $d=100\pm 22~\mathrm{nm}$ for ``deep'' clusters.

\newpage
\section{Dipole moment direction}
In this work, without loss of generality, we focus on $(100)$-oriented diamond. The $(100)$-surface cut forms the air-diamond interface at $z=0$. The $z$-axis of our coordinate system is aligned with the $[100]$ direction. On the other hand, the $x$-axis is chosen to be aligned with the $[00\bar{1}]$ direction, while the $y$-axis is chosen to be aligned with the $[010]$ direction. 
For each SiV$^{-}$ center, the three-fold symmetry axis is randomly assigned from the four equivalent directions $[111]$, $[1\bar{1}\bar{1}]$, $[\bar{1}\bar{1}1]$, and $[\bar{1}1\bar{1}]$. These four directions should be projected onto our coordinate system:
\begin{align}
    &[111]: \left(-\frac{1}{\sqrt{3}}, \frac{1}{\sqrt{3}}, \frac{1}{\sqrt{3}} \right), \\
    &[1\bar{1}\bar{1}]: \left(\frac{1}{\sqrt{3}}, -\frac{1}{\sqrt{3}}, \frac{1}{\sqrt{3}} \right), \\
    &[\bar{1}\bar{1}1]: \left( -\frac{1}{\sqrt{3}}, -\frac{1}{\sqrt{3}}, -\frac{1}{\sqrt{3}} \right), \\
    &[\bar{1}1\bar{1}]: \left( \frac{1}{\sqrt{3}}, \frac{1}{\sqrt{3}}, -\frac{1}{\sqrt{3}} \right).
\end{align}
The axis of each SiV$^{-}$ center is sampled randomly from the above four directions. The transition dipole orientation $\bm{\mu}_{i}$ is parallel to this axis \cite{Hepp2014, Rogers2014a}. The amplitude of dipole moment $\bm{\mu}_{i}$ is determined using the radiative decay rate inside homogeneous environment:
\begin{equation}
    \Gamma^{\text{homo}} = \frac{n \mu_{i}^{2} \omega_{0}^{3}}{3\pi \hbar \epsilon_{0} c_{0}^{3}},
\end{equation}
where $n=2.4$ denotes the refractive index of diamond. 

\newpage
\section{Monte Carlo simulation: convergence test}
\begin{figure}[h]
\centering
\includegraphics[width=1\linewidth]{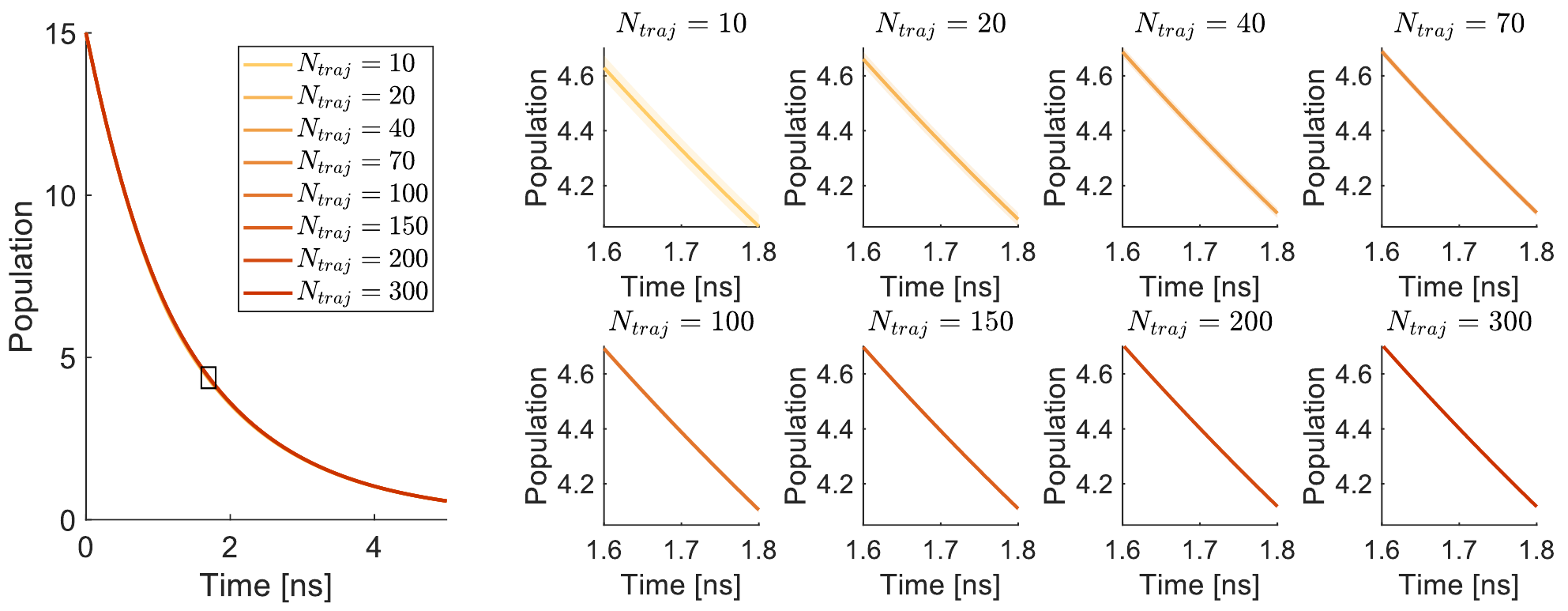}
\caption{Convergence test of Monte Carlo simulation, using $N=15$ SiV$^{-}$ centers. The averaged population curves differ very little for different $N_\text{traj}$ values. In the right panel, the standard error curves have been visualized using shaded areas. }
\label{fig:convergence_test}
\end{figure}

In this part we intend to test the convergence of our Monte Carlo simulation. The simulation setup remains the same as the ones shown in Figure 3 of the main text. Specifically, we use $N=15$ SiV$^{-}$ centers, with depth $d=100\pm 22$ nm. The quantum efficiency remains $\eta_{Q}=60\%$. The number of trajectories included in the Monte Carlo simulation is denoted as $N_\text{traj}$. We focus on the averaged value of the excited state population. The corresponding curves are plotted in Figure~\ref{fig:convergence_test}. In order to understand how the accuracy of our estimation changes when increasing the number of trajectories, we further calculate the standard error of the obtained trajectories, calculated using
\begin{equation}
    \text{SE} = \frac{\sigma}{\sqrt{N_\text{traj}}} = \frac{1}{\sqrt{N_\text{traj}}} \cdot \sqrt{\frac{1}{N_\text{traj} - 1} \sum_{i=1}^{N_\text{traj}} (x_{i} - \overline{x})^{2}}.
\end{equation}
The results are displayed in the right panel of Figure~\ref{fig:convergence_test}. For a given $N_\text{traj}$, the corresponding $\pm \text{SE}$ range is marked using the shaded area. Note that we choose to zoom-in and inspect the $1.6\text{ns}<t<1.8\text{ns}$ range for clarity. For $N_\text{traj}=300$, the standard error is smaller than $0.0064~P_\text{exc}$, indicating that using $300$ trajectories would be enough. 

\newpage
\section{Near-field interaction: suppressed superradiance}
\begin{figure}[h]
\centering
\includegraphics[width=0.5\linewidth]{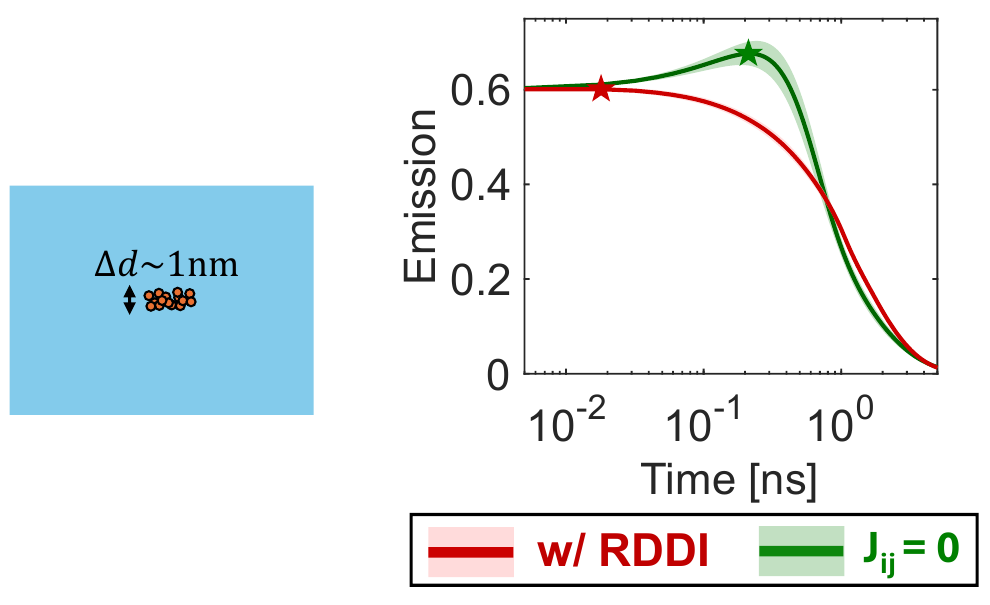}
\caption{Suppressed superradiance of a dense SiV$^{-}$ cluster. The density distribution in $z$ direction has a standard deviation of $1$ nm. When considering near-field dipole-dipole interaction (displayed as the red curve), the photon emission rate only shows a negligible burst. }
\label{fig:1nm}
\end{figure}

As mentioned in the main text, in R. Dicke's original paper \cite{Dicke1954} he believed that a small cluster of emitters will show strong superradiance. It was later pointed out that, in a small cluster with high density, the presence of strong near-field dipole-dipole interaction actually suppresses superradiance \cite{Friedberg1972}. 
In this part we aim to provide an extra data point to demonstrate that higher density does not lead to stronger superradiance, due to the near-field interaction. We consider $N=15$ SiV$^{-}$ centers inside homogeneous environment, so the effect of air-diamond interface can be disregarded. The positions of SiV$^{-}$ centers inside the $xy$ plane are sampled uniformly inside a circular region (diameter $D=60$ nm), while the $z$ coordinates are sampled from Gaussian distribution with a standard deviation of $\Delta d = 1$ nm. Such narrow depth distribution requires nanometer-precision depth control, which can be achieved using delta-doping technique \cite{Ohno2012_APL}. The density of SiV$^{-}$ centers can be estimated as
\begin{equation}
    \frac{N}{V} \approx \frac{N}{\sqrt{2\pi} \Delta d \cdot \pi (D/2)^2} \approx 2.1\times 10^{18}~\text{cm}^{-3}, 
\end{equation}
As a contrast, a typical cold atom gas has a density of $10^{11}\sim 10^{13}~\text{cm}^{-3}$ \cite{Araujo2016}, which is orders of magnitude lower compared to what can be achieved in solid state systems. Other parameters remain the same as Figure 3 in the main text. 
The collective emission dynamics of such a high-density SiV$^{-}$ ensemble is shown in Figure~\ref{fig:1nm}. Similar to the results displayed in the main text, the mean value (indicated by the solid lines) and the standard deviation (indicated by the shaded areas) are obtained based on $300$ different cluster configurations. It can be concluded that higher density does not lead to larger superradiant burst. On the other hand, when the coherent part $J_{ij}$ of dipole-dipole interaction is set as $0$, an apparent superradiant burst appears. This justifies that in a dense cluster, the strong near-field interaction suppresses superradiance. 

\newpage
\section{CUDA implementation}
In this part, we provide implementation details of the program used to calculate the PLE spectra, shown in Figure 4d of the main text. 
As mentioned in the main text, the PLE spectrum is measured by exciting the SiV$^{-}$ centers with a tunable laser. In order to predict the PLE spectrum using the simulation technique, we also have to sweep the laser detuning $\Delta = \omega_{0} - \omega_{d}$. For each $\Delta$ the equations of motion are solved until reaching steady-state. The incident wave shows the spatial profile of a Gaussian beam, whose waist at $z=0$ can be calculated by $w_{0}=\frac{\lambda_{0}}{\pi\cdot \text{NA}}\approx 427$ nm when using an objective lens with $\text{NA}=0.55$. At the point with depth $z$ and radius $r$, the electric field is defined by
\begin{equation}
    E(r, z) = E_{0} \cdot \frac{w_{0}}{w(z)}\cdot \exp\left( -\frac{r^2}{w(z)^{2}} \right) \cdot 
    \exp\left[ -ikz - \frac{ikr^{2}}{2R(z)} + i\zeta(z) \right],
\end{equation}
where $z_{R}=\frac{n\pi w_{0}^{2}}{\lambda_{0}}$ denotes the Rayleigh length inside diamond, $\zeta(z)=\arctan(\frac{z}{z_{R}})$ denotes the Gouy phase; $w(z)=w_{0} \sqrt{1+(\frac{z}{z_{R}})^{2}}$ denotes the beam radius, and $R(z) = z\cdot \left[1+(\frac{z_{R}}{z})^{2}\right]$ denotes the wavefront radius \cite{Novotny2006}. 

Notice that the same procedure has to be carried out for all different clusters involved in our Monte Carlo simulation. This procedure requires a very large number of ODE solving to obtain faithful averaged results, which quickly becomes computationally expensive for large $N$. 
This issue requires us to examine the computational complexity more carefully. Under our cumulant expansion framework, only cumulants up to the second-order are involved in the ODEs. Take $\langle \hat{\sigma}^{21}_i \hat{\sigma}^{12}_j \rangle$ as an example: 
\begin{align}
\frac{d}{dt}\langle \hat{\sigma}^{21}_i \hat{\sigma}^{12}_j \rangle = 
&\left[i(\delta_i - \delta_j) - \frac{\Gamma_{ii} + \Gamma_{i}^{\text{nrad}}}{2} - \frac{\Gamma_{jj} + \Gamma_{j}^{\text{nrad}}}{2}\right] \langle \hat{\sigma}^{21}_i \hat{\sigma}^{12}_j \rangle \nonumber \\
&+ \frac{\Gamma_{ji}}{2} \left[4 \langle \hat{\sigma}^{22}_i \hat{\sigma}^{22}_j \rangle - \langle \hat{\sigma}^{22}_i \rangle - \langle \hat{\sigma}^{22}_j \rangle \right]  +{iJ_{ji}( \hat{\sigma}^{22}_j - \hat{\sigma}^{22}_i ) } \nonumber \\
&+ \sum_{k \neq i,j} \left(iJ_{ki} - \frac{\Gamma_{ki}}{2}\right) \left[\langle \hat{\sigma}^{21}_k \hat{\sigma}^{12}_j \rangle - 2\langle \hat{\sigma}^{21}_k \hat{\sigma}^{22}_i \hat{\sigma}^{12}_j \rangle \right] \nonumber \\
&+ \sum_{k \neq i,j} \left(-iJ_{jk} - \frac{\Gamma_{jk}}{2}\right) \left[\langle \hat{\sigma}^{12}_k \hat{\sigma}^{21}_i \rangle - 2\langle \hat{\sigma}^{12}_k \hat{\sigma}^{21}_i \hat{\sigma}^{22}_j \rangle \right] \nonumber \\
&+ \frac{i\Omega_i^\ast}{2} \left(\langle \hat{\sigma}^{12}_j \rangle - 2\langle \hat{\sigma}^{22}_i \hat{\sigma}^{12}_j \rangle \right) - \frac{i\Omega_j}{2} \left(\langle \hat{\sigma}^{21}_i \rangle - 2\langle \hat{\sigma}^{21}_i \hat{\sigma}^{22}_j \rangle \right).
\end{align}
Therefore the number of ODEs scales as $O(N^2)$. However, for each time-step, the naive implementation needs to iterate over $i$, $j$ and $k$ indices explicitly, leading to $O(N^3)$ complexity. Fortunately, the complexity issue can be alleviated by taking advantage of the massive number of CUDA cores contained in modern GPUs. Specifically, we map different $(i, j)$ pairs onto different CUDA threads, since calculating the time derivatives for different $(i, j)$ pairs are independent tasks, thus can be completed in parallel. Each CUDA thread will now only have to iterate over $k$ index. 

In the CUDA-accelerated version, the ODEs are solved using the classic 4-th order Runge-Kutta method. Specifically, denote the time-step as $h$. At each time-step, the update rules for $\frac{dy}{dt}=f(t, y)$ follows:
\begin{align}
    &k_{1} = f(t, y), \nonumber \\
    &k_{2} = f(t+\frac{h}{2}, y+\frac{k_{1}h}{2}), \nonumber \\
    &k_{3} = f(t+\frac{h}{2}, y+\frac{k_{2}h}{2}), \nonumber \\
    &k_{4} = f(t+h, y+k_{3}h), \nonumber \\
    &y \rightarrow y+\frac{h}{6}(k_{1}+2k_{2}+2k_{3}+k_{4}), \nonumber \\
    &t \rightarrow t+h.
\end{align}
In our implementation we fix the time-step $h$ as $h=\min\{ \frac{0.2}{\max_{i,j}\|J_{ij}\|}, \frac{0.2}{\max_{i,j}\|\Gamma_{ij}\|} \}$, in order to ensure that the interval $h$ is small enough to capture the fast dynamics. 
For $N=15$ SiV$^{-}$ centers, the naive MATLAB implementation (runs on Intel(R) Xeon(R) CPU E5-2650 v2 @2.60GHz) typically takes more than $400$ seconds to solve a single set of ODEs. As a comparison, our CUDA-accelerated implementation, which runs on a GPU (NVIDIA Titan Xp graphic card, with 12GB global memory), takes around $11$ seconds on average. This leads to $>36\times$ acceleration, making the calculation of PLE spectrum computationally feasible. 

\newpage
\section{Correlation function \& emission spectrum}
In the main text, it has been mentioned that calculating the emission spectrum requires calculating the correlation function $C_{ij}(\omega) = \int_{-\infty}^{\infty} d\tau \left\langle \hat{\sigma}_{i}^{21}(t) \hat{\sigma}_{j}^{12}(t+\tau) \right\rangle e^{i\omega \tau}$. In this part we provide details on how the correlation function can be calculated with the help of quantum regression theorem \cite{Carmichael1993}. 

Suppose that we aim to calculate correlation function $\langle \sigma_k^{21}(t) \sigma_i^{12}(t+\tau) \rangle$. Assume that at time $t$ the system has already reached its steady state, the starting time $t$ can be set as $0$. 
The time-evolution of $\textcolor{red}{\langle \sigma_k^{21}(0) \sigma_i^{12}(\tau) \rangle}$ reads:
\begin{align}
\frac{d}{d\tau} \textcolor{red}{\langle \sigma_k^{21}(0) \sigma_i^{12}(\tau) \rangle} &=
\left[-i(\Delta + \delta_i) - \frac{\Gamma_{ii}}{2}\right]  \textcolor{red}{\langle \sigma_k^{21}(0) \sigma_i^{12}(\tau) \rangle}
+ \frac{i\Omega_i}{2} \left( 2 \textcolor{blue}{\langle \sigma_k^{21}(0) \sigma_i^{22}(\tau) \rangle} - \langle \sigma_k^{21}(0)\rangle  \right) \notag \\
&+ \sum_{j \neq i} \left( iJ_{ij} + \frac{\Gamma_{ij}}{2} \right) 
\cdot \left( 2\langle \sigma_k^{21}(0) \sigma_j^{12}(\tau) \sigma_i^{22}(\tau) \rangle -  \textcolor{red}{\langle \sigma_k^{21}(0) \sigma_j^{12}(\tau) \rangle} \right).
\end{align}
By applying the second-order cumulant expansion 
\begin{align}
\langle \sigma_k^{21}(0) \sigma_j^{12}(\tau) \sigma_i^{22}(\tau) \rangle 
&\approx \langle \sigma_k^{21}(0) \rangle \langle \sigma_j^{12}(0) \sigma_i^{22}(0) \rangle 
+ \langle \sigma_j^{12}(0) \rangle  \textcolor{blue}{\langle \sigma_k^{21}(0) \sigma_i^{22}(\tau) \rangle} \notag \\
&\quad + \langle \sigma_i^{22}(0) \rangle  \textcolor{red}{ \langle \sigma_k^{21}(0) \sigma_j^{12}(\tau) \rangle }
- 2 \langle \sigma_k^{21}(0) \rangle \langle \sigma_j^{12}(0) \rangle \langle \sigma_i^{22}(0) \rangle,
\end{align}
we obtain the first equation:
\begin{align}
\frac{d}{d\tau} \textcolor{red}{\langle \sigma_k^{21}(0) \sigma_i^{12}(\tau) \rangle}
&= \left[-i(\Delta + \delta_i) - \frac{\Gamma_{ii}}{2}\right]
   \textcolor{red}{\langle \sigma_k^{21}(0) \sigma_i^{12}(\tau) \rangle}
 + \frac{i\Omega_i}{2} \left( 2 \textcolor{blue}{ \langle \sigma_k^{21}(0) \sigma_i^{22}(\tau) \rangle}
   -  \langle \sigma_k^{21}(0)\rangle \right) \notag\\
&\quad + \sum_{j \neq i} 2\!\left( iJ_{ij} + \frac{\Gamma_{ij}}{2} \right)
 \Bigg[
   \langle \sigma_k^{21}(0)\rangle
   \Big( \langle \sigma_j^{12}(0) \sigma_i^{22}(0) \rangle
   - 2\langle \sigma_j^{12}(0) \rangle \langle \sigma_i^{22}(0) \rangle \Big) \notag\\
&\quad
 + \langle \sigma_j^{12}(0) \rangle \,
   \textcolor{blue}{ \langle \sigma_k^{21}(0) \sigma_i^{22}(\tau) \rangle}
 \;+\;
 \left( \langle \sigma_i^{22}(0) \rangle - \frac{1}{2} \right)
   \textcolor{red}{ \langle \sigma_k^{21}(0) \sigma_j^{12}(\tau) \rangle }
 \Bigg].
\end{align}
The time-evolution of $\textcolor{orange}{\langle \sigma_k^{21}(0) \sigma_i^{21}(\tau) \rangle }$ reads:
\begin{align}
\frac{d}{d\tau} \textcolor{orange}{\langle \sigma_k^{21}(0) \sigma_i^{21}(\tau) \rangle }
&= \left[ i (\Delta + \delta_i) - \frac{\Gamma_{ii}}{2} \right] 
   \textcolor{orange}{\langle \sigma_k^{21}(0) \sigma_i^{21}(\tau) \rangle }
    - \frac{i \Omega_i^\ast}{2} \big( 2 \textcolor{blue}{\langle \sigma_k^{21}(0) \sigma_i^{22}(\tau) \rangle }-  \langle \sigma_k^{21}(0)\rangle \big) \notag \\
&\quad + \sum_{j \neq i} 
    \left( -i J_{ji} + \frac{\Gamma_{ji}}{2} \right) 
    \cdot \Big( 
        2 \langle \sigma_k^{21}(0) \sigma_j^{21}(\tau) \sigma_i^{22}(\tau) \rangle 
        -\textcolor{orange}{ \langle \sigma_k^{21}(0) \sigma_j^{21}(\tau) \rangle }
    \Big).
\end{align}
By applying the second-order cumulant expansion 
\begin{align}
\langle \sigma_k^{21}(0) \sigma_j^{21}(\tau) \sigma_i^{22}(\tau) \rangle 
&\approx \langle \sigma_k^{21}(0) \rangle \langle \sigma_j^{21}(0) \sigma_i^{22}(0) \rangle 
+ \langle \sigma_j^{21}(0) \rangle \textcolor{blue}{ \langle \sigma_k^{21}(0) \sigma_i^{22}(\tau) \rangle } \notag \\
&\quad + \langle \sigma_i^{22}(0) \rangle \textcolor{orange}{\langle \sigma_k^{21}(0) \sigma_j^{21}(\tau) \rangle }
- 2 \langle \sigma_k^{21}(0) \rangle \langle \sigma_j^{21}(0) \rangle \langle \sigma_i^{22}(0) \rangle,
\end{align}
we obtain the second equation:
\begin{align}
\frac{d}{d\tau} \textcolor{orange}{ \langle \sigma_k^{21}(0) \sigma_i^{21}(\tau) \rangle}
&= \left[ i (\Delta + \delta_i) - \frac{\Gamma_{ii}}{2} \right]
   \textcolor{orange}{\langle \sigma_k^{21}(0) \sigma_i^{21}(\tau) \rangle }
   - \frac{i \Omega_i^\ast}{2}
     \left( 2 \textcolor{blue}{ \langle \sigma_k^{21}(0) \sigma_i^{22}(\tau) \rangle }
            -  \langle \sigma_k^{21}(0)\rangle \right) \notag\\
&\quad + \sum_{j \neq i} 2\left( -i J_{ji} + \frac{\Gamma_{ji}}{2} \right)
\Bigg[
   \langle \sigma_k^{21}(0) \rangle
   \Big( \langle \sigma_j^{21}(0) \sigma_i^{22}(0) \rangle
        - 2 \langle \sigma_j^{21}(0) \rangle \langle \sigma_i^{22}(0) \rangle \Big) \notag\\
&\quad
 + \langle \sigma_j^{21}(0) \rangle \,
   \textcolor{blue}{ \langle \sigma_k^{21}(0) \sigma_i^{22}(\tau) \rangle }
 \;+\;
 \left( \langle \sigma_i^{22}(0) \rangle - \frac{1}{2} \right)
   \textcolor{orange}{\langle \sigma_k^{21}(0) \sigma_j^{21}(\tau) \rangle}
\Bigg].
\end{align}
Similarly, the time-evolution of $\textcolor{blue}{\langle \sigma_k^{21}(0) \sigma_i^{22}(\tau) \rangle }$ reads:
\begin{align}
\frac{d}{d\tau} &\textcolor{blue}{\langle \sigma_k^{21}(0) \sigma_i^{22}(\tau) \rangle }
= -\Gamma_{ii} \textcolor{blue}{\langle \sigma_k^{21}(0) \sigma_i^{22}(\tau) \rangle }
    - \frac{1}{2} \big( i \Omega_i \textcolor{orange}{\langle \sigma_k^{21}(0) \sigma_i^{21}(\tau) \rangle }
    - i \Omega_i^\ast \textcolor{red}{\langle \sigma_k^{21}(0) \sigma_i^{12}(\tau) \rangle }\big) \notag \\
& + \sum_{j \neq i} \Bigg[ \left( i J_{ji} - \frac{\Gamma_{ji}}{2} \right) 
    \langle \sigma_k^{21}(0) \sigma_j^{21}(\tau) \sigma_i^{12}(\tau) \rangle 
    + \left( -i J_{ij} - \frac{\Gamma_{ij}}{2} \right) 
    \langle \sigma_k^{21}(0) \sigma_j^{12}(\tau) \sigma_i^{21}(\tau) \rangle \Bigg].
\end{align}
By applying the second-order cumulant expansion 
\begin{align}
\langle \sigma_k^{21}(0) \sigma_j^{21}(\tau) \sigma_i^{12}(\tau) \rangle 
&\approx \langle \sigma_k^{21}(0) \rangle \langle \sigma_j^{21}(0) \sigma_i^{12}(0) \rangle 
+ \langle \sigma_j^{21}(0) \rangle\textcolor{red}{ \langle \sigma_k^{21}(0) \sigma_i^{12}(\tau) \rangle } \notag \\
&\quad + \langle \sigma_i^{12}(0) \rangle \textcolor{orange}{ \langle \sigma_k^{21}(0) \sigma_j^{21}(\tau) \rangle }
- 2 \langle \sigma_k^{21}(0) \rangle \langle \sigma_j^{21}(0) \rangle \langle \sigma_i^{12}(0) \rangle,
\end{align}
\begin{align}
\langle \sigma_k^{21}(0) \sigma_j^{12}(\tau) \sigma_i^{21}(\tau) \rangle 
&\approx \langle \sigma_k^{21}(0) \rangle \langle \sigma_j^{12}(0) \sigma_i^{21}(0) \rangle 
+  \langle \sigma_j^{12}(0) \rangle \textcolor{orange}{ \langle \sigma_k^{21}(0) \sigma_i^{21}(\tau) \rangle }  \notag \\
&\quad + \langle \sigma_i^{21}(0) \rangle\textcolor{red}{ \langle \sigma_k^{21}(0) \sigma_j^{12}(\tau) \rangle }
- 2 \langle \sigma_k^{21}(0) \rangle \langle \sigma_j^{12}(0) \rangle \langle \sigma_i^{21}(0) \rangle,
\end{align}
we obtain the third equation:
\begin{align}
\frac{d}{d\tau} \textcolor{blue}{ \langle \sigma_k^{21}(0) \sigma_i^{22}(\tau) \rangle }
&= -\Gamma_{ii} \textcolor{blue}{ \langle \sigma_k^{21}(0) \sigma_i^{22}(\tau) \rangle }
   - \frac{1}{2}\Big( i \Omega_i \textcolor{orange}{\langle \sigma_k^{21}(0) \sigma_i^{21}(\tau) \rangle }
   - i \Omega_i^\ast \textcolor{red}{\langle \sigma_k^{21}(0) \sigma_i^{12}(\tau) \rangle } \Big) \notag\\
&\quad + \sum_{j \neq i} \left( i J_{ji} - \frac{\Gamma_{ji}}{2} \right)
\Bigg[
   \langle \sigma_k^{21}(0) \rangle
   \Big( \langle \sigma_j^{21}(0) \sigma_i^{12}(0) \rangle
       - 2\langle \sigma_j^{21}(0) \rangle \langle \sigma_i^{12}(0) \rangle \Big) \notag\\
&\qquad\qquad\quad
 + \langle \sigma_j^{21}(0) \rangle \,
   \textcolor{red}{ \langle \sigma_k^{21}(0) \sigma_i^{12}(\tau) \rangle }
 \;+\;
   \langle \sigma_i^{12}(0) \rangle \, \textcolor{orange}{ \langle \sigma_k^{21}(0) \sigma_j^{21}(\tau) \rangle }
\Bigg] \notag\\
&\quad + \sum_{j \neq i}  \left( -i J_{ij} - \frac{\Gamma_{ij}}{2} \right)
\Bigg[
   \langle \sigma_k^{21}(0) \rangle
   \Big( \langle \sigma_j^{12}(0) \sigma_i^{21}(0) \rangle
       - 2 \langle \sigma_j^{12}(0) \rangle \langle \sigma_i^{21}(0) \rangle \Big) \notag\\
&\qquad\qquad\quad
 + \langle \sigma_i^{21}(0) \rangle \, \textcolor{red}{ \langle \sigma_k^{21}(0) \sigma_j^{12}(\tau) \rangle }
 \;+\;
   \langle \sigma_j^{12}(0) \rangle \, \textcolor{orange}{ \langle \sigma_k^{21}(0) \sigma_i^{21}(\tau) \rangle }
\Bigg].
\end{align}
After solving the above equations numerically, the results are Fourier transformed to obtain $C_{ij}(\omega)$, which is then used to calculate the emission spectrum.  

\newpage
\section{Single-excitation states}
In this part we explain in details how we obtained the single-excitation eigenstates shown in Figure 4c, 4e, 4f. More specifically, consider $N$ coupled TLSs. Following \cite{Garcia2017a}, to obtain the single-excitation eigenstate, we can describe the system using a non-Hermitian effective Hamiltonian
\begin{equation}
\mathcal{H}_{\text{eff}} =\sum_{i}\hbar (\omega_{0} + \delta_{i} - i\frac{\Gamma_{ii} }{2})\cdot \hat{\sigma}_{i}^{21} \hat{\sigma}_{i}^{12} + \sum_{j\neq i} \hbar (J_{ij}-i\frac{\Gamma_{ij}}{2}) \cdot \hat{\sigma}_{i}^{21} \hat{\sigma}_{j}^{12}. 
\label{eq:H_eff}
\end{equation}
The single-excitation quantum state can be expressed as 
\begin{align}
    | \Psi(t) \rangle = \sum_{i} b_{i}(t) |e_{i}\rangle 
    =\sum_{i} b_{i}(t) \cdot\hat{\sigma}_{i}^{21} |g_{1}, g_{2}, ..., g_{N}\rangle, 
\end{align}
where $|g_{1}, g_{2}, ..., g_{N}\rangle$ represents the state where all TLSs are at ground state, $|e_{i}\rangle$ represents the state where only the $i$-th TLS is excited, and $b_{i}$ are the corresponding coefficients. Based on Schr$\ddot{\text{o}}$dinger equation, the time-evolution of coefficients $b_{i}(t)$ can be derived as
\begin{equation}
    \frac{\partial b_{i}}{\partial t} = \left[ -i(\omega_{0} + \delta_{i}) - \frac{\Gamma_{ii}}{2} \right] b_{i} 
    + \sum_{j\neq i} (-iJ_{ij} - \frac{\Gamma_{ij}}{2}) b_{j}. 
\end{equation}
The resonance frequencies of single-excitation eigenstates can thus be extracted through diagonalizing the following matrix:
\begin{equation}
    \langle e_i | \mathcal{H}_{\text{eff}} | e_j \rangle = \hbar (\omega_{0} + \delta_{i}-i\frac{\Gamma_{ii}}{2})\cdot \delta_{ij} + \hbar (J_{ij}-i\frac{\Gamma_{ij}}{2})\cdot(1-\delta_{ij}),
\end{equation}
where $\delta_{ij}$ represents Kronecker delta. 

\newpage
\section{Estimation of skewness}
As explained in the main text, the skewness value has been used to quantify the asymmetry of a density-of-states (DOS) distribution. In order to estimate the skewness value $\mathbb{E}\left[\left(\frac{X-\mu}{\sigma}\right)^{3} \right]$, we obtain the sample skewness by calculating the adjusted Fisher-Pearson standardized moment coefficient. More specifically, for a set of data points $\{ x_{i} \}$ ($i=1,2,..., n$), the skewness can be estimated by calculating \cite{Doane2011, SKEW_excel}
\begin{equation}
    G_{1} = \frac{n}{(n-1)(n-2)}\cdot \sum_{i} \left( \frac{x_{i} - \overline{x}}{\sigma} \right)^{3}, 
\end{equation}
where $\overline{x} = \frac{1}{n} \sum_{i} x_{i}$ stands for the mean value, and $\sigma=\sqrt{\frac{1}{n-1} \sum_{i} (x_{i} - \overline{x})^{2}}$ denotes the unbiased estimation of standard deviation. In Figure 4e and 4f of the main text, the data points $x_{i}$ correspond to the calculated eigenfrequencies of single-excitation states. 


\end{document}